\documentclass[twocolumn,english,pre,showpacs]{revtex4}
\usepackage{ae,aecompl}
\usepackage[T1]{fontenc}
\usepackage[latin9]{inputenc}
\usepackage{array}
\usepackage{amsmath}
\usepackage{graphicx}
\usepackage{amssymb}
\usepackage{esint}

\makeatletter

\providecommand{\tabularnewline}{\\}

\@ifundefined{textcolor}{}
{%
 \definecolor{BLACK}{gray}{0}
 \definecolor{WHITE}{gray}{1}
 \definecolor{RED}{rgb}{1,0,0}
 \definecolor{GREEN}{rgb}{0,1,0}
 \definecolor{BLUE}{rgb}{0,0,1}
 \definecolor{CYAN}{cmyk}{1,0,0,0}
 \definecolor{MAGENTA}{cmyk}{0,1,0,0}
 \definecolor{YELLOW}{cmyk}{0,0,1,0}
 }

\newcommand{\C}[1]{{}} 
\def\Tr{\mathop{\operator@font Tr}\nolimits}
\def\diag{\mathop{\operator@font diag}\nolimits}
\def\EpK{\mathop{\operator@font K}\nolimits}
\def\EpE{\mathop{\operator@font E}\nolimits}
\def\EpPi{\mathop{\operator@font \Pi}\nolimits}
\def\artanh{\mathop{\operator@font artanh}\nolimits}

\makeatother

\usepackage{babel}

\begin{document}
\global\long\def\Lp{L_{\parallel}}
\global\long\def\Ls{L_{\perp}}
\global\long\def\bc{\beta_{\mathrm{c}}}
\global\long\def\Tc{T_{\mathrm{c}}}
\global\long\def\Kc{K_{\mathrm{c}}}
\global\long\def\para{\parallel}
\global\long\def\bH{\beta\mathcal{H}}
\global\long\def\B{\mathrm{b}}
\global\long\def\mat#1{\mathbf{#1}}
\global\long\def\Se{1\mathrm{d}}
\global\long\def\Szb{2\mathrm{d}_{\mathrm{b}}}
\global\long\def\See{1{+}1\mathrm{d}}
\global\long\def\Sz{2\mathrm{d}}
\global\long\def\Sdb{3\mathrm{d}_{\mathrm{b}}}
\global\long\def\Sze{2{+}1\mathrm{d}}
\global\long\def\Sd{3\mathrm{d}}

\title{Non--equilibrium phase transition in an exactly solvable\\
 driven Ising model with friction}

\author{Alfred Hucht}

\affiliation{Fakultät für Physik und CeNIDE, Universität Duisburg-Essen, D-47048
Duisburg}

\pacs{05.50.+q, 68.35.Rh, 04.20.Jb}

\date{\today}
\begin{abstract}
A driven Ising model with friction due to magnetic correlations has
recently been proposed by Kadau \emph{et al}. {[}Phys. Rev. Lett.
\textbf{101}, 137205 (2008){]}. The non--equilibrium phase transition
present in this system is investigated in detail using analytical
methods as well as Monte Carlo simulations. In the limit of high driving
velocities $v$ the model shows mean field behavior due to dimensional
reduction and can be solved exactly for various geometries. The simulations
are performed with three different single spin flip rates: the common
Metropolis and Glauber rates as well as a \emph{multiplicative} rate.
Due to the non--equilibrium nature of the model all rates lead to
different critical temperatures at $v>0$, while the exact solution
matches the multiplicative rate. Finally, the cross--over from Ising
to mean field behavior as function of velocity and system size is
analysed in one and two dimensions. 
\end{abstract}
\maketitle

\section{Introduction}

\begin{figure}
\begin{centering}
\includegraphics[width=8cm]{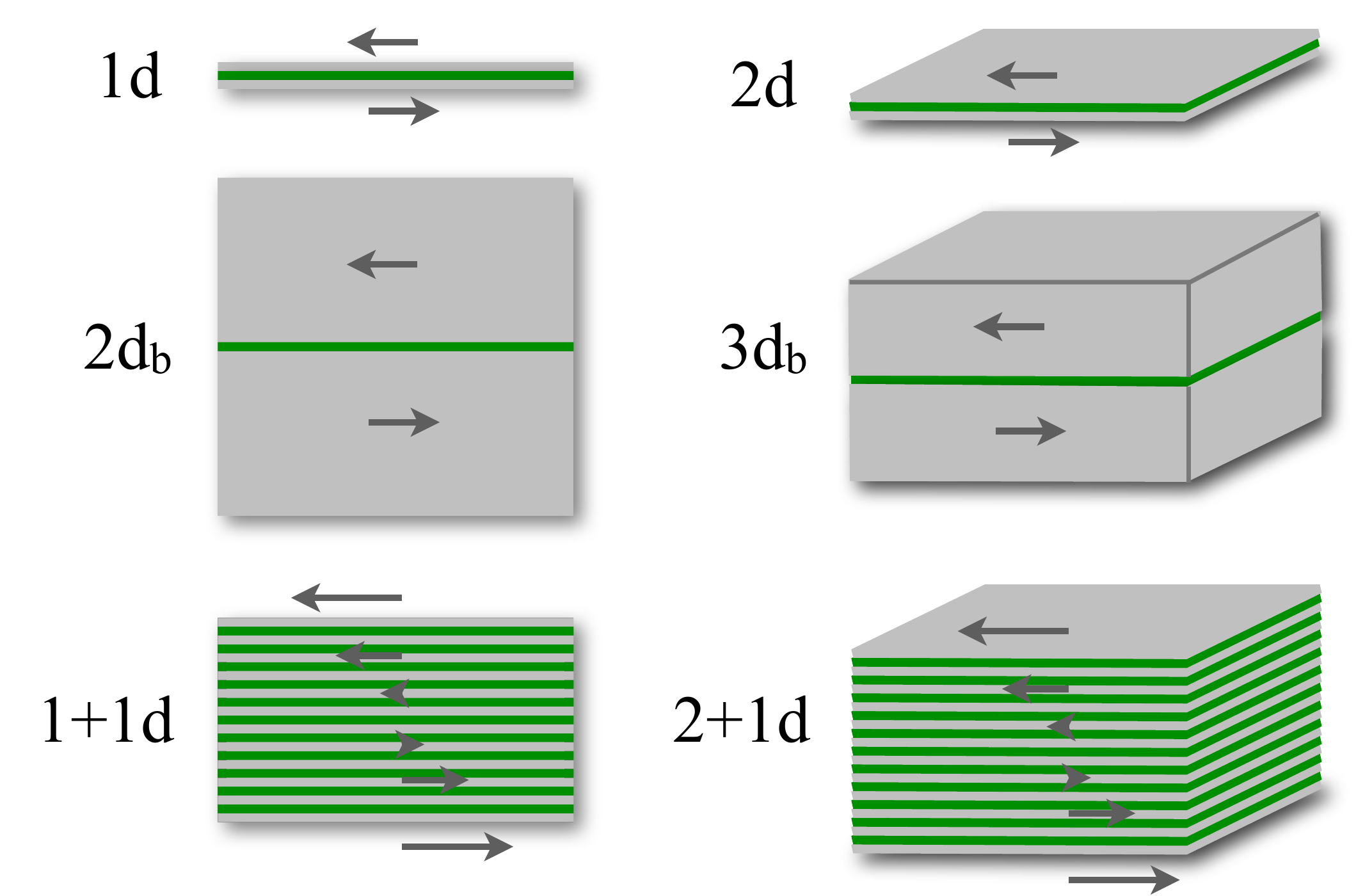}
\par\end{centering}

\caption{(Color online) Overview of the geometries considered in this work.
The grey regions are the magnetic systems, while the green (dark)
regions are the moving boundaries. The arrows indicate the motion
of the subsystems.\label{fig:Geometries}}

\end{figure}
Magnetic contributions to friction due to spin correlations have attracted
increasing interest in recent years. One interesting aspect is the
energy dissipation due to spin waves in magnetic force microscopy,
where magnetic structures are investigated by moving a magnetic tip
over a surface \citep{FuscoWolfNowak08,MagieraBrendelWolfNowak09,MagieraWolfBrendelNowak09}.
On the other hand, magnetic friction is also present in bulk magnetic
systems which are in close proximity. In this context, Kadau \emph{et
al.~}\citep{KadauHuchtWolf08} proposed a simple model for magnetic
friction mediated solely by spin degrees of freedom. In this model
an Ising spin system is moved over a second spin system with constant
velocity $v$ along a boundary. This permanent perturbation drives
the system to a steady state far away from equilibrium, leading to
a permanent energy flow from the boundary to the heat bath.

This problem can be analyzed for several different geometries in one,
two and three dimensions, as shown in Fig.~\ref{fig:Geometries}:
Besides the original problem of two half-infinite two dimensional
systems moving along the one dimensional boundary, denoted $\Szb$
in the following, we will consider the homogeneous cases $\Se$ and
$\Sz$ where all spins are at the boundary, as well as the experimentally
relevant three dimensional case $\Sdb$. Additionally, we will extend
the analysis to sheared systems in two \citep{ChanLin90,CorberiGonnellaLamura98,CirilloGonnellaSaracco05}
and three \citep{ImaekaKawasaki84} dimensions, denoted $\See$ and
$\Sze$. These systems are experimentally accessible within the framework
of shear flow in binary liquid mixtures (for a review, see \citep{Onuki97}),
though with conserved order parameter, while we deal with a non--conserved
order parameter.

This model has some similarities to the \emph{driven lattice gas}
(DLG) proposed by Katz \emph{et al}.~\citep{KatzLebowitzSpohn83}
(see \citep{SchmittmannZia95} for a review), where a system is driven
out of equilibrium by an applied field which favors the motion of
particles in one direction. We will discuss these similarities throughout
this work. 

The paper is organized as follows: In the first part we will introduce
the model and geometries and present, in the second part, an exact
solution of the model in the limit of high driving velocities $v\rightarrow\infty$,
which will be checked numerically in the last part using Monte Carlo
simulations. There we will also investigate the case of finite velocities
$v$.

\section{Model \label{sec:Model}}

Let us start with the simplest case denoted $\Se$ in Fig.~\ref{fig:Geometries}
and consider two Ising chains with spin variables $\sigma=\pm1$,
nearest neighbor coupling $K=\beta J$ ($\beta=1/k_{\mathrm{B}}T$
and we set $k_{\mathrm{B}}=1$) and $\Lp$ sites each, interacting
with boundary coupling $K_{\B}=\beta J_{\B}$ and moving along each
other with relative velocity $v$. In the Monte Carlo simulation the
upper system is moved $v$ times by one lattice constant $a_{0}$
with respect to the lower system during each random sequential Monte
Carlo sweep (MCS). As one MCS corresponds to a typical spin relaxation
time $t_{0}=\mathcal{O}(10^{-8}\mathrm{s})$ \citep{Stearns86} and
$a_{0}=\mathcal{O}(10^{-10}\mathrm{m})$, the velocity $v$ is given
in natural units $a_{0}/t_{0}=\mathcal{O}(1\mathrm{cm/s})$ (we will
set $a_{0}=t_{0}=1$ in the following). 

\begin{figure}
\begin{centering}
\includegraphics[clip,scale=0.7]{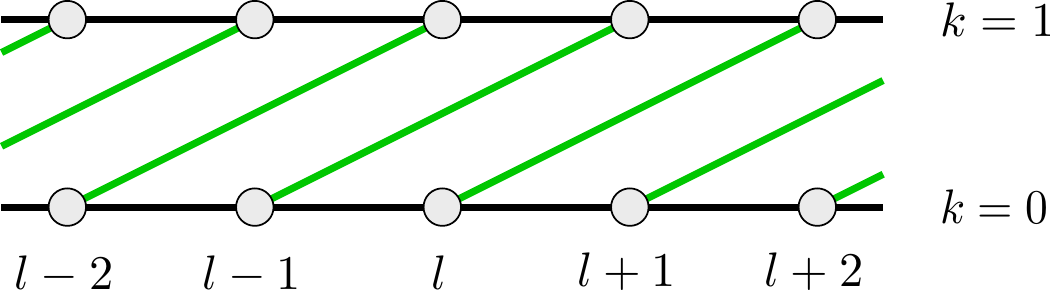}
\par\end{centering}

\caption{(Color online) Sketch of geometry $\Se$ after $\Delta=2$ moves.
Spin $\sigma_{0,l}$ interacts with spin $\sigma_{1,l+2}$ with coupling
$J_{\B}$ (green (gray) lines), while all other couplings are $J$
(black lines). \label{fig:system_1d}}

\end{figure}
To simplify the implementation, instead of moving the upper part of
the lattice with respect to the lower part we reorder the couplings
at the boundary with time. This procedure is analogous to the Lees--Edwards
or \emph{moving boundary condition} in molecular dynamics simulations
of fluids \citep{AllenTildesley87} and leads to a system as shown
in Fig.~\ref{fig:system_1d}. Assuming periodic boundary conditions
(PBC) $\sigma_{k,l}\equiv\sigma_{k,l\bmod\Lp}$ in the parallel direction,
the time dependent Hamiltonian reads \begin{equation}
\bH(t)=-K\sum_{k=0}^{1\vphantom{\Lp}}\sum_{l=1}^{\Lp}\sigma_{k,l}\sigma_{k,l+1}-K_{\B}\sum_{l=1}^{\Lp}\sigma_{0,l}\sigma_{1,l+\Delta(t)}\label{eq:H_1d}\end{equation}
with the time dependent displacement \begin{equation}
\Delta(t)=vt.\label{eq:shift}\end{equation}

\begin{figure}[b]
\begin{centering}
\includegraphics[bb=0bp 0bp 313bp 223bp,clip,scale=0.7]{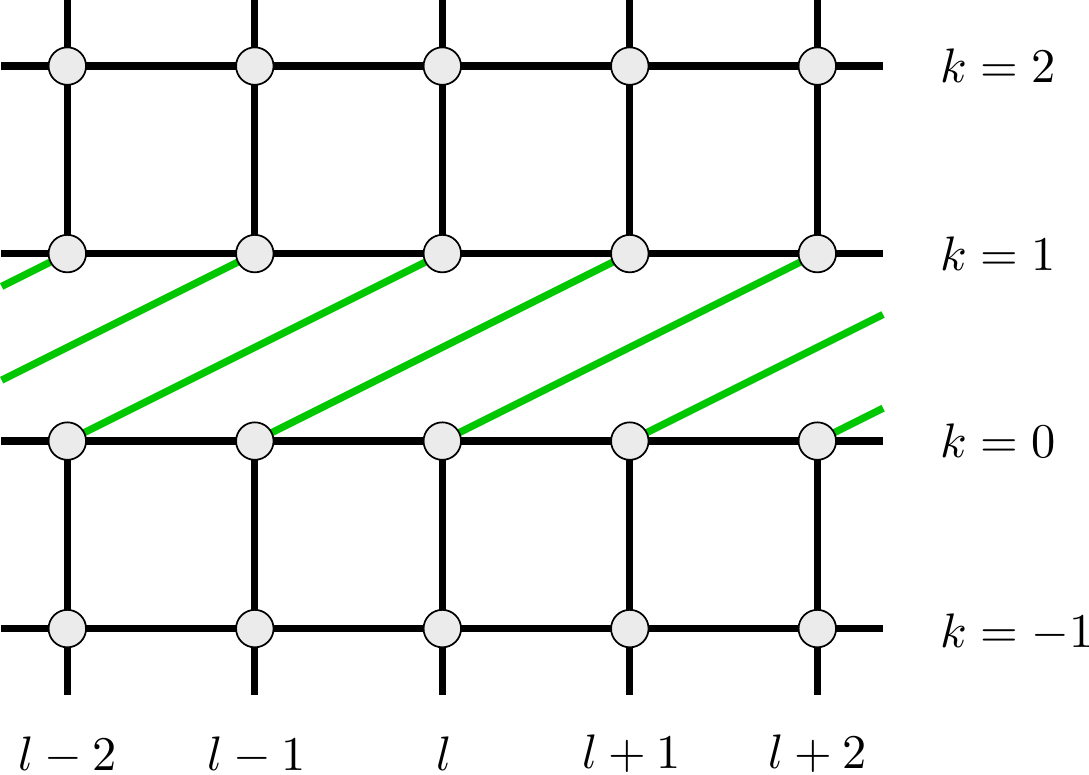}
\par\end{centering}

\caption{(Color online) Sketch of geometry $\Szb$ after $\Delta=2$ moves.
\label{fig:system_2db}}

\end{figure}
The second geometry considered in this work is the $\Szb$ case shown
in Fig.~\ref{fig:system_2db}, which already was investigated by
Kadau \emph{et al}.~\citep{KadauHuchtWolf08}. Here we have a square
lattice with $\Lp\times\Ls$ sites and periodic boundary conditions
in both directions, i.\,e., $\sigma_{k,l}\equiv\sigma_{k\bmod\Ls,l\bmod\Lp}$.
Note that especially $\sigma_{\Ls,l}\equiv\sigma_{0,l}$. The Hamiltonian
of this system becomes \begin{equation}
\bH(t)=-\sum_{k=1}^{\Ls\vphantom{\Lp}}\sum_{l=1}^{\Lp}K\sigma_{k,l}\sigma_{k,l+1}+K_{\perp,k}\sigma_{k,l}\sigma_{k+1,l+\Delta_{k}(t)}\label{eq:H_2d}\end{equation}
with $\Delta_{k}(t)\equiv0$ and $K_{\perp,k}=K$ for all rows except
row $k=0$, where the couplings to row $k=1$ are shifted with constant
velocity $\Delta_{0}(t)\equiv\Delta(t)=vt.$ The coupling $K_{\perp,0}\equiv K_{\B}$
across the boundary is allowed to be different from $K$. For $v=0$
and $J_{\B}=J=1$ this system simplifies to the $2d$ Ising model
in equilibrium, which was solved exactly by \citet{Onsager44} and
shows a continuous phase transition at \begin{equation}
T_{\mathrm{c,eq}}=\frac{2}{\log(1+\sqrt{2})}=2.2691853\ldots\,.\label{eq:Tc_Ising}\end{equation}
Note that both systems are translationally invariant in $\para$ direction
under the transformation $l\rightarrow l+1$ and obey reflection symmetry
at the boundary under $k\rightarrow1-k$.

\section{Exact solution at high velocities \label{sec:Exact_Solution}}

\begin{figure}
\begin{centering}
\includegraphics[scale=0.7]{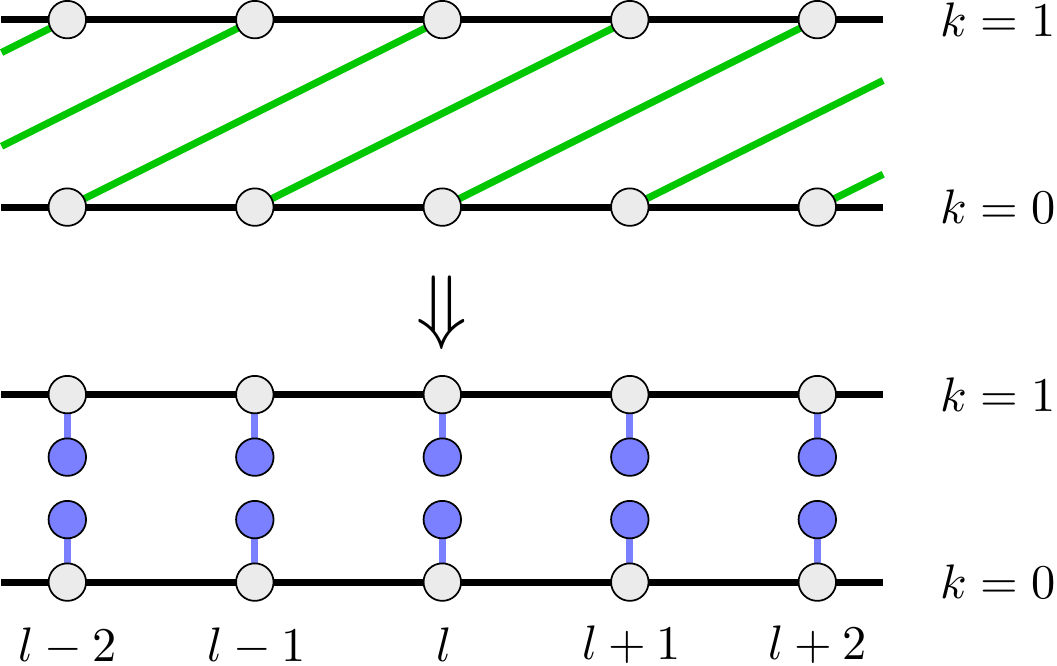}
\par\end{centering}

\caption{(Color online) Mapping of the $\Se$ driven system, shown for $\Delta=2$,
on two disconnected $1d$ systems with fluctuating fields. \label{fig:mapping_1d}}

\end{figure}

In Ref.~\citep{KadauHuchtWolf08} it was shown that for high velocities
$v\gg1$ the properties of the $\Szb$ system become independent of
$v$. This can be understood as follows: In the limit $v\rightarrow\infty$
the interaction $K_{\B}\sigma_{0,l}\sigma_{1,l+\Delta(t)}$ $ $across
the driven boundary becomes uncorrelated, as, in the Monte Carlo simulations,
at large $v$ the spin $\sigma_{1,l+\Delta(t)}$ is different in every
trial step and can, for simplicity, be a randomly chosen spin $\sigma_{1,\mathrm{rnd}}$
from row $1$. Note that this simplification was checked within the
simulations and indeed gave the same results, enabling us to perform
simulations \emph{at} $v=\infty$. Thus the boundary coupling can
be replaced by the action of a fluctuating boundary field $\mu$,
e.g., \begin{equation}
\sigma_{0,l}\sigma_{1,l+\Delta(t)}\rightarrow\sigma_{0,l}\sigma_{1,\mathrm{rnd}}\rightarrow\sigma_{0,l}\mu_{0,l},\label{eq:mapping}\end{equation}
with stochastic variables $\mu_{kl}=\pm1$ ($k=0,1$) under the constraint
$\langle\mu_{kl}\rangle=\langle\sigma_{kl}\rangle=m_{\B}$, where
$m_{\B}$ denotes the magnetization at the driven boundary. Here we
used the translation symmetry $\langle\sigma_{kl}\rangle=m_{k}$ and
the reflection symmetry at the boundary, $m_{k}=m_{1-k}$. In Fig.~\ref{fig:mapping_1d}
this mapping of the driven system onto a system with fluctuating boundary
fields is illustrated for the $\Se$ case. The next step will be to
map the fluctuating fields onto static fields by integrating out the
degrees of freedom $\mu_{kl}$.

\subsection{Ising model in a fluctuating field}

Consider a general Ising model with arbitrary couplings $K_{ij}$
in a static external field $h_{i}^{\mathrm{ext}}$ and additional
fluctuating fields of strength $k_{i}$ (note the factor $\beta$
in all field quantities) \begin{equation}
\bH_{\mu}=-\sum_{i<j}K_{ij}\sigma_{i}\sigma_{j}-\sum_{i}(h_{i}^{\mathrm{ext}}+k_{i}\mu_{i})\sigma_{i}\label{eq:H_general_tilde}\end{equation}
 where the $\mu_{i}=\pm1$ are stochastic variables at site $i$ with
given average \begin{equation}
\langle\mu_{i}\rangle=m_{i}.\label{eq:m_i_condition}\end{equation}
As this condition is given \emph{a priori}, averages containing $\mu_{i}$
can be calculated using the trace formula \begin{equation}
\Tr_{\mu}f(\mu_{i})=\sum_{\mu_{i}=\pm1}f(\mu_{i})p_{i}(\mu_{i})\label{eq:Tr_mu}\end{equation}
with the probability distribution $p_{i}(\mu_{i})=(1+\mu_{i}m_{i})/2$,
as then \[
\langle\mu_{i}\rangle=\Tr_{\mu}\mu_{i}=\sum_{\mu_{i}=\pm1}\mu_{i}\, p_{i}(\mu_{i})=m_{i}\]
 as assumed. With the decomposition \begin{equation}
\bH_{\mu}=\bH_{0}-\sum_{i}k_{i}\mu_{i}\sigma_{i}\label{eq:H_tilde_decom}\end{equation}
the degrees of freedom $\mu$ in the partition function $\mathcal{Z}$
can be traced out, \begin{eqnarray}
\mathcal{Z} & = & \Tr_{\sigma\mu}e^{-\bH_{\mu}}\:=\:\Tr_{\sigma}e^{-\beta\mathcal{H}_{0}}\Tr_{\mu}\prod_{i}e^{k_{i}\mu_{i}\sigma_{i}}\nonumber \\
 & = & \Tr_{\sigma}e^{-\beta\mathcal{H}_{0}}\prod_{i}\sum_{\mu_{i}=\pm1}e^{k_{i}\mu_{i}\sigma_{i}}p_{i}(\mu_{i})\nonumber \\
 & = & \prod_{i}\cosh k_{i}\Tr_{\sigma}e^{-\beta\mathcal{H}_{0}}\prod_{i}\left[1+\sigma_{i}m_{i}\tanh k_{i}\right],\label{eq:Z_tilde}\end{eqnarray}
where we used the fact that $\sigma_{i}=\pm1$. 

On the other hand, the Hamiltonian of the equilibrium Ising model
without fluctuating fields in a static field $h_{i}$ can be written
as \begin{equation}
\bH_{\mathrm{eq}}=-\sum_{i<j}K_{ij}\sigma_{i}\sigma_{j}-\sum_{i}h_{i}\sigma_{i}=\bH_{0}-\sum_{i}b_{i}\sigma_{i}\label{eq:H_general}\end{equation}
with $\mathcal{H}_{0}$ from Eq.~(\ref{eq:H_tilde_decom}), if we
let $b_{i}=h_{i}-h_{i}^{\mathrm{ext}}$. The partition function of
this model clearly fulfills \begin{eqnarray}
\mathcal{Z}_{\mathrm{eq}} & = & \Tr_{\sigma}e^{-\bH_{\mathrm{eq}}}\:=\:\Tr_{\sigma}e^{-\bH_{0}}\prod_{i}e^{b_{i}\sigma_{i}}\nonumber \\
 & = & \prod_{i}\cosh b_{i}\Tr_{\sigma}e^{-\bH_{0}}\prod_{i}\left[1+\sigma_{i}\tanh b_{i}\right].\label{eq:Z_eq}\end{eqnarray}
Comparing Eqs.~(\ref{eq:Z_tilde}) and (\ref{eq:Z_eq}), we conclude
that under the condition\begin{equation}
\tanh b_{i}=m_{i}\tanh k_{i}\label{eq:condition}\end{equation}
the partition function $\mathcal{Z}$ can be expressed in terms of
$\mathcal{Z}_{\mathrm{eq}}$, \begin{equation}
\mathcal{Z}=\left.\prod_{i}\frac{\cosh k_{i}}{\cosh b_{i}}\,\mathcal{Z}_{\mathrm{eq}}\right|_{\mathrm{Eq.\,(\ref{eq:condition})}}.\label{eq:Z_condition}\end{equation}

To summarize, the coupling with strength $k_{i}$ to fluctuating fields
$\mu_{i}=\pm1$ with given average $\langle\mu_{i}\rangle=m_{i}$
can be written as coupling to static effective fields $b_{i}$ with
strength given by Eq.~(\ref{eq:condition}). In the next section
we will use this mapping to exactly solve the driven Ising model for
high velocities $v\rightarrow\infty$.

\subsection{Application to the driven Ising model}

The general condition Eq.~(\ref{eq:condition}) for the effective
static fields $b_{i}$ simplifies for the systems considered in this
work: As all boundary spins are equivalent, $m_{i}=m_{\B}$, with
coupling $k_{i}=K_{\B}$, leading to a uniform effective field $h_{\B}=\artanh(m_{\B}\tanh K_{\B})$
at the boundary, as we assume no additional external fields, $h_{i}^{\mathrm{ext}}=0$.
Inserting this into the equilibrium expression for the boundary magnetization
$m_{\B,\mathrm{eq}}(K,h_{\B})=\partial\ln\mathcal{Z}_{\mathrm{eq}}/\partial h_{\B}$,
we end with the self--consistence condition \begin{equation}
m_{\B,\mathrm{eq}}[K,\artanh(m_{\B}\tanh K_{\B})]=m_{\B}\label{eq:sc}\end{equation}
for the non--equilibrium order parameter $m_{\B}$. 

As $1=\partial m_{\B,\mathrm{eq}}/\partial m_{\B}|_{m_{\B}=0}$ at
criticality, we obtain a very useful connection between the reduced
zero field boundary susceptibility of the equilibrium model $\chi_{\B,\mathrm{eq}}^{(0)}(K)=\partial m_{\B,\mathrm{eq}}/\partial h_{\B}|_{h_{\B}=0}$
and the critical temperature $\Tc$ of the driven system by expanding
Eq.~(\ref{eq:sc}) to first order around $m_{\B}=0$, namely \begin{equation}
\chi_{\B,\mathrm{eq}}^{(0)}(K_{\mathrm{c}})\,\tanh K_{\B,\mathrm{c}}=1.\label{eq:chi_Tc}\end{equation}
In the following we will apply these results to the one and two dimensional
model introduced in Section~\ref{sec:Model}.

\subsection{1d case\label{sub:Exact_1d}}

The effective Hamiltonian of the system $\Se$ in a fluctuating field
reads

\begin{equation}
\bH=-\sum_{l=1}^{\Lp}K\sigma_{l}\sigma_{l+1}+(h^{\mathrm{ext}}+K_{\B}\mu_{l})\sigma_{l}.\label{eq:H_1d_eff}\end{equation}
Applying the self--consistence condition Eq.~(\ref{eq:sc}) to the
well known expression for the equilibrium magnetization of the $1d$
Ising model \citep[cf.][]{Baxter82} \begin{equation}
m_{\mathrm{eq}}(K,h)=\frac{\sinh h}{\sqrt{e^{-4K}+\sinh^{2}h}}\label{eq:mag1d_eq}\end{equation}
we obtain the zero field magnetization of the $\Se$ driven system
in the ordered phase for velocity $v\rightarrow\infty$, \begin{equation}
m(K,K_{\B})=\sqrt{\frac{\cosh2K_{\B}-\coth2K}{\cosh2K_{\B}-1}}\label{eq:mag1d}\end{equation}
 with critical temperature fulfilling \begin{equation}
e^{2K_{\mathrm{c}}}\tanh K_{\B,\mathrm{c}}=1,\label{eq:Tc_1d}\end{equation}
as $\chi_{\mathrm{eq}}^{(0)}(K)=e^{2K}$ in this case. Interestingly,
Eq.~(\ref{eq:mag1d}) is equal to the spontaneous surface magnetization
of the $2d$ equilibrium Ising model \citep[Chapter VI, Eq. 5.20]{McCoyWu73}
if we identify $K$ and $K_{\B}$ with the couplings $\para$ and
$\perp$ to the surface, and consequently has the identical critical
temperature $\Tc$. For the special case $K=K_{\B}$ this gives the
well known value from Eq.~(\ref{eq:Tc_Ising}). However, we regard
this equality as coincidence without deeper meaning, as Eq.~(\ref{eq:mag1d})
is solution of a simple quadratic equation with small integer coefficients
when written in the natural variables. Nevertheless, we checked this
identity in the $\Sz$ case and found that we do \emph{not} get the
surface magnetization of the $3d$ system by the same procedure, as
the critical temperature is $\Tc\approx4.058$ (Eq.~(\ref{eq:Tc_2d}))
instead of the correct value $\Tc=4.511424(53)$ \citep{FerrenbergLandau91,ItoHukushimaOgawaOzeki00}. 

To calculate other quantities we use the transfer matrix (TM) formulation:
the TM of the $1d$ equilibrium Ising model reads \citep[cf.][]{Baxter82}
\begin{equation}
\mat T_{\mathrm{eq}}=\left(\begin{array}{cc}
e^{K+h} & e^{-K}\\
e^{-K} & e^{K-h}\end{array}\right)\label{eq:TM_eq}\end{equation}
and the partition function of a periodic system with $\Lp$ spins
can be expressed as \begin{equation}
\mathcal{Z}_{\mathrm{eq}}=\Tr\mat T_{\mathrm{eq}}^{\Lp}.\label{eq:Z_TM_eq}\end{equation}
Using Eq.~(\ref{eq:Z_condition}) and the conditions Eq.~(\ref{eq:condition})
we can write \begin{equation}
\mathcal{Z}=\Tr\mat T^{\Lp}\label{eq:Z_TM}\end{equation}
with the TM (we set $h^{\mathrm{ext}}=0$ from now on)\begin{equation}
\mat T=\left.\frac{\cosh K_{\B}}{\cosh h}\,\mat T_{\mathrm{eq}}\right|_{\tanh h=m\tanh K_{\B}},\label{eq:TM_1}\end{equation}
which can be written as \begin{equation}
\mat T=\cosh K_{\B}\left(\begin{array}{cc}
e^{K}\left(1+\sin\psi\right) & e^{-K}\cos\psi\\
e^{-K}\cos\psi & e^{K}\left(1-\sin\psi\right)\end{array}\right)\label{eq:TM}\end{equation}
using \begin{equation}
\sin\psi=m\tanh K_{\B}.\label{eq:psi}\end{equation}
The angle $\psi$ decreases from $\psi=\pi/2$ at $T=0$ to $\psi=0$
at $T\ge\Tc$. The eigenvalues $\lambda_{\mu}$ of $\mat T$ fulfill
\begin{equation}
\mat T|t_{\mu}\rangle=\lambda_{\mu}|t_{\mu}\rangle\label{eq:TM_es}\end{equation}
and are given by\begin{equation}
\lambda_{0,1}=\left\{ \begin{array}{lll}
{\displaystyle e^{K\pm K_{\B}}} &  & T\leq\Tc\\
{\displaystyle \cosh K_{\B}(e^{K}\pm e^{-K})} &  & T\geq\Tc\end{array}\right.\label{eq:lambda}\end{equation}
where $\lambda_{0}$ denotes the larger eigenvalue dominant in the
thermodynamic limit. Note that in this limit the analog to the free
energy density \begin{equation}
f=-\frac{1}{\beta}\log\lambda_{0}=-(J+J_{\B})\label{eq:free}\end{equation}
of the driven system is simply a constant in the ordered phase $T\leq\Tc$
\footnote{TM calculations for stripes of width $\Ls>1$ show that this is only
the case for $\Ls=1$.%
}. Nevertheless, we can calculate physical quantities within this TM
notation using expectation values, as the whole information of the
half-infinite system is contained in the normalized eigenvectors \begin{equation}
|t_{0}\rangle=\left(\begin{array}{c}
\cos\phi\\
\sin\phi\end{array}\right),\;|t_{1}\rangle=\left(\begin{array}{c}
-\sin\phi\\
\hphantom{-}\cos\phi\end{array}\right)\label{eq:t_mu}\end{equation}
with $\cos2\phi=m$. Using the normalized TM $\hat{\mat T}=\mat T/\lambda_{0}$
and the Pauli matrix $\mat M=\diag(1,-1)$, the magnetization, Eq.~(\ref{eq:mag1d}),
can be expressed as\begin{equation}
m=\langle t_{0}|\mat M|t_{0}\rangle,\label{eq:m_again}\end{equation}
 while the correlation function in $\para$ direction becomes \begin{eqnarray}
g_{\para}(n) & = & \langle\sigma_{l}\sigma_{l+n}\rangle-\langle\sigma_{l}\rangle\langle\sigma_{l+n}\rangle\nonumber \\
 & = & \langle t_{0}|\mat M\hat{\mat T}^{n}\mat M|t_{0}\rangle-\langle t_{0}|\mat M|t_{0}\rangle^{2}\nonumber \\
 & = & \lambda_{1}^{n}\lambda_{0}^{-n}\langle t_{0}|\mat M|t_{1}\rangle^{2},\label{eq:g_para_def}\end{eqnarray}
as $\mat T^{n}=\sum_{\mu}\lambda_{\mu}^{n}|t_{\mu}\rangle\langle t_{\mu}|$.
We get the result \begin{equation}
g_{\para}(n)=\left\{ \begin{array}{lll}
{\displaystyle (1-m^{2})e^{-2nK_{\B}}} &  & T\leq\Tc\\
{\displaystyle \tanh^{n}K} &  & T\geq\Tc\end{array}\right.,\label{eq:g_para}\end{equation}
 \C{\[
[[[\;\langle\sigma_{l}\sigma_{l+n}\rangle_{<}=1-(1-e^{-2nK_{\B}})(1-m^{2})\;]]]\]
}leading to the inverse correlation length \begin{equation}
\xi_{\para}^{-1}=\log\frac{\lambda_{0}}{\lambda_{1}}=\left\{ \begin{array}{lll}
{\displaystyle 2K_{\B}} &  & T\leq\Tc\\
{\displaystyle \log\coth K} &  & T\geq\Tc\end{array}\right..\label{eq:xi}\end{equation}
Note that $\xi_{\para}$ does not diverge at the critical point, a
feature which would lead to a correlation length exponent $\nu=0$.
In Section~\ref{sec:Monte-Carlo-Simulations} we will argue that
in finite systems the spin fluctuations are not only mediated by the
spins $\sigma_{i}$ but also by the self consistent field $m$ which
fluctuates at finite $\Lp$, an effect which vanishes in the exact
solution, as $\Lp\to\infty$. 

\begin{figure}
\begin{centering}
\includegraphics[width=0.9\columnwidth]{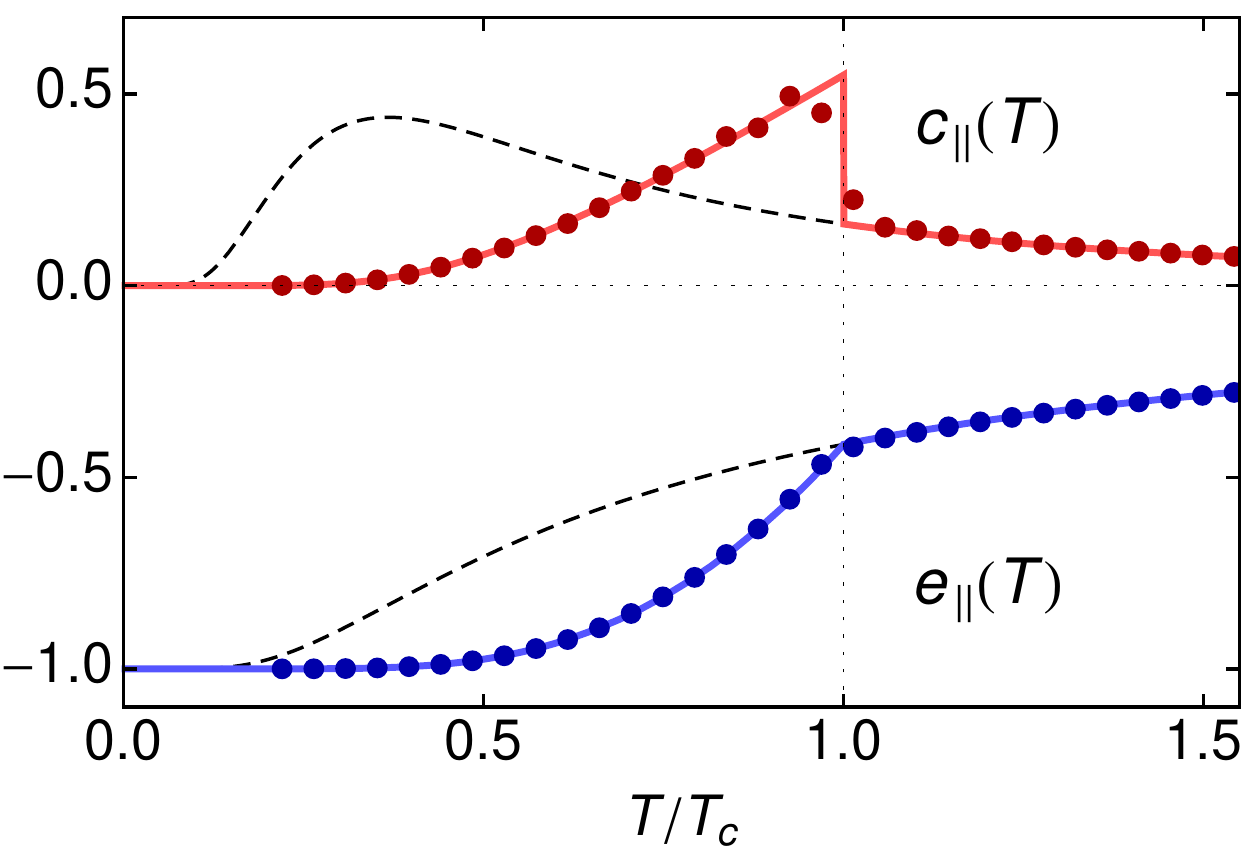}
\par\end{centering}

\caption{(Color online) Internal energy $e_{\para}(T)$, Eq.~(\ref{eq:e_para}),
and specific heat $c_{\para}(T)$, Eq.~(\ref{eq:c_para}), of the
$\Se$ driven system at $v\to\text{\ensuremath{\infty}}$. The points
are MC results for $\Lp=2^{11}$, and the dashed lines are results
for the one dimensional Ising model in equilibrium.}

\end{figure}
From the nearest neighbor correlation function we can calculate the
internal energy $e_{\para}=-J\langle\sigma_{l}\sigma_{l+1}\rangle$
in $\para$ direction \begin{equation}
e_{\para}=\left\{ \begin{array}{lll}
{\displaystyle \frac{Je^{-2K-K_{\B}}}{\sinh2K\sinh K_{\B}}-1} &  & T\leq\Tc\\
\\{\displaystyle -J\tanh K} &  & T\geq\Tc\end{array}\right.\label{eq:e_para}\end{equation}
 as well as the specific heat $c_{\para}=\partial e_{\para}/\partial T$
in $\para$ direction \begin{equation}
c_{\para}=\left\{ \begin{array}{lll}
{\displaystyle \frac{2K^{2}}{\sinh^{2}K}(\coth K_{\B}-1)} &  & T<\Tc\\
\\{\displaystyle \frac{K^{2}}{\cosh^{2}K}} &  & T>\Tc\end{array}\right..\label{eq:c_para}\end{equation}
On the other hand, the internal energy  in $\perp$ direction is simply
given by \begin{equation}
e_{\perp}=-J_{\B}m^{2}\label{eq:e_perp}\end{equation}
as the related spins are uncorrelated. 

Now we turn to dynamical properties of this system under a concrete
MC Glauber dynamics (see Sec.~\ref{sub:An-integrable-algorithm}
for details) and calculate the spin flip acceptance rate $A=\langle p_{\mathrm{flip}}\rangle$
and the energy dissipation rate $P=\partial E/\partial t$: Let $\langle\zeta_{\ell}\,\zeta\,\zeta_{r}\rangle$
denote the probability of picking a spin $\sigma$ with direction
$\zeta=\uparrow,\downarrow$ and left and right neighbors $\sigma_{\ell,r}$
with direction $\zeta_{\ell,r}$. These probabilities can be calculated
using the matrices $\mat P_{\uparrow}=\diag(1,0)$ and $\mat P_{\downarrow}=\diag(0,1)$,
e.\,g., $\langle\uparrow\uparrow\downarrow\rangle=\langle t_{0}|\mat P_{\uparrow}\hat{\mat T}\mat P_{\uparrow}\hat{\mat T}\mat P_{\downarrow}|t_{0}\rangle$.
As the third coupling partner $\mu$ of spin $\sigma$, with direction
$\zeta_{\mu}$, is uncorrelated at infinite velocity, the probability
of a particular spin configuration becomes \begin{equation}
\langle\zeta_{\ell}\,\zeta\,\zeta_{r}\rangle\langle\zeta_{\mu}\rangle=\langle t_{0}|\mat P_{\zeta_{\ell}}\hat{\mat T}\mat P_{\zeta}\hat{\mat T}\mat P_{\zeta_{r}}|t_{0}\rangle\langle t_{0}|\mat P_{\zeta_{\mu}}|t_{0}\rangle.\label{eq:p_conf}\end{equation}
 The spin flip probability of a given configuration is $p_{\mathrm{flip}}(\Delta E)$,
with $\Delta E=\Delta E_{1}+\Delta E_{2}=2J\sigma(\sigma_{\ell}+\sigma_{r})+2J_{\B}\sigma\mu$,
and $A$ becomes the sum over all $2^{4}$ possible cases \begin{equation}
A=\sum_{\zeta_{\ell},\zeta,\zeta_{r},\zeta_{\mu}=\uparrow,\downarrow}p_{\mathrm{flip}}(\Delta E)\langle\zeta_{\ell}\,\zeta\,\zeta_{r}\rangle\langle\zeta_{\mu}\rangle,\label{eq:A_def}\end{equation}
which can be written as \begin{eqnarray}
A & = & \sum_{\zeta_{\ell},\zeta,\zeta_{r}=\uparrow,\downarrow}p_{\mathrm{flip}}^{*}(\Delta E_{1})\langle\zeta_{\ell}\,\zeta\,\zeta_{r}\rangle\sum_{\zeta_{\mu}=\uparrow,\downarrow}p_{\mathrm{flip}}^{*}(\Delta E_{2})\langle\zeta_{\mu}\rangle\nonumber \\
 & = & \sum_{\zeta=\uparrow,\downarrow}X_{\zeta}\left(e^{-2K_{\B}}\langle\zeta\rangle+\langle\bar{\zeta}\rangle\right)\label{eq:A_mult}\end{eqnarray}
for the multiplicative rate $p_{\mathrm{flip}}^{*}(\Delta E)=p_{\mathrm{flip}}^{*}(\Delta E_{1})p_{\mathrm{flip}}^{*}(\Delta E_{2})$
introduced in Sec.~\ref{sub:An-integrable-algorithm}, Eq.~(\ref{eq:p_flip_FR}),
using the abbreviation\begin{eqnarray}
X_{\zeta} & = & \sum_{\zeta_{\ell},\zeta_{r}=\uparrow,\downarrow}p_{\mathrm{flip}}^{*}(\Delta E_{1})\langle\zeta_{\ell}\,\zeta\,\zeta_{r}\rangle\nonumber \\
 & = & e^{-4K}\langle\zeta\zeta\zeta\rangle+2e^{-2K}\langle\zeta\zeta\bar{\zeta}\rangle+\langle\bar{\zeta}\zeta\bar{\zeta}\rangle.\label{eq:X_zeta}\end{eqnarray}
Note that the two terms in Eq.~(\ref{eq:A_mult}) are equal and the
acceptance rate is independent of spin $\zeta$ because $m$ is stationary.
The resulting acceptance rate becomes \begin{equation}
A=\left\{ \begin{array}{lll}
{\displaystyle \frac{\cosh(K+K_{\B})-\sinh(K-K_{\B})}{4e^{2(K+K_{\B})}\sinh K\cosh^{2}K\sinh K_{\B}}} &  & T\leq\Tc\\
\\e^{-K_{\B}}\cosh K_{\B}(1-\tanh K)^{2} &  & T\geq\Tc\end{array}\right.,\label{eq:A_1d}\end{equation}
which simplifies for $J=J_{\B}$ to \begin{equation}
A=\left\{ \begin{array}{lll}
{\displaystyle \frac{e^{-4K}\coth2K}{\sinh2K}} &  & T\leq\Tc\\
\\{\displaystyle \frac{e^{-3K}}{\cosh K}} &  & T\geq\Tc\end{array}\right..\label{eq:A_1d1}\end{equation}

\begin{figure}
\begin{centering}
\includegraphics[width=0.9\columnwidth]{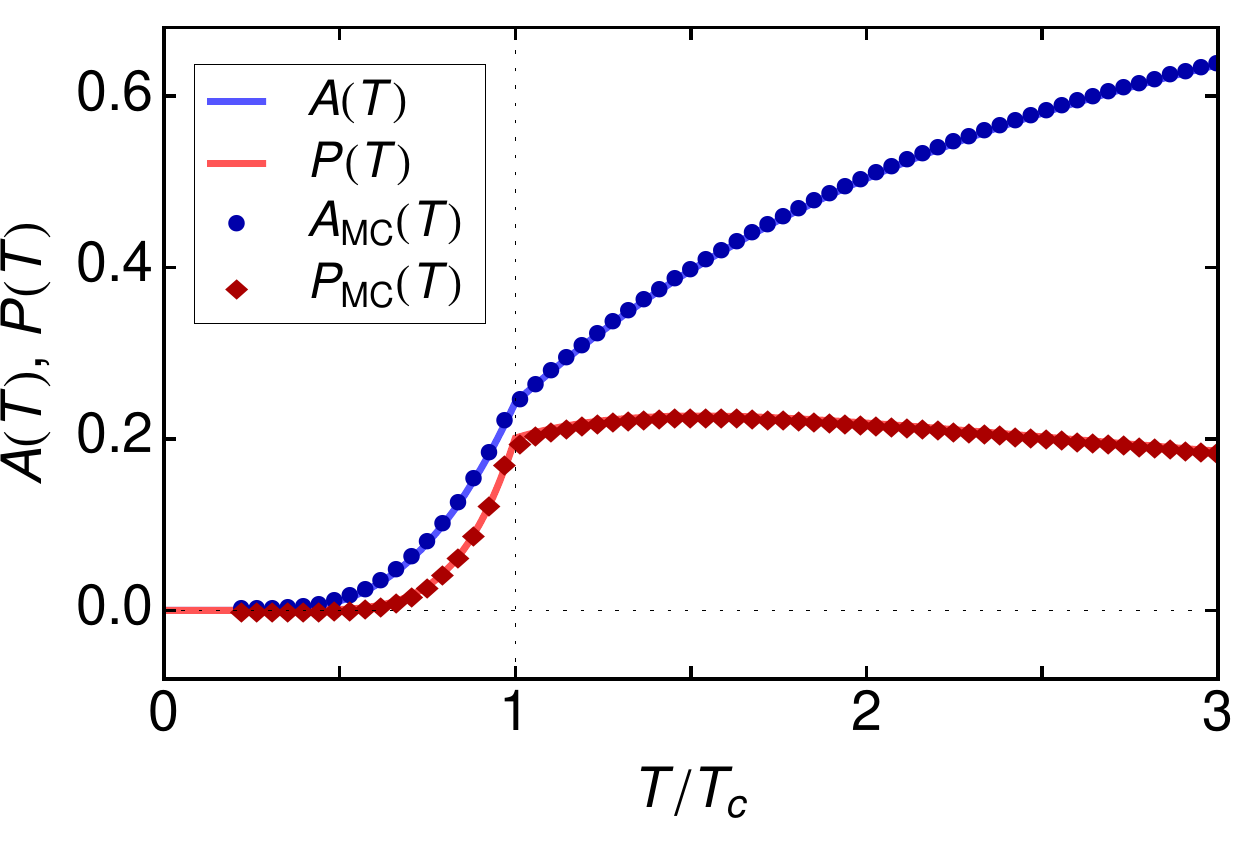}
\par\end{centering}

\caption{(Color online) Spin flip probability $A$, Eq.~(\ref{eq:A_1d1}),
and energy dissipation rate $P$, Eq.~(\ref{eq:P_1d_1}), versus
reduced temperature $T/\Tc$ for the $\Se$ system at $v\to\infty$,
together with MC data for $\Lp=2^{11}$. \label{fig:A_and_P_1d}}

\end{figure}
The calculation of the energy dissipation rate $P$ per spin is very
similar to the acceptance rate $A$ (Eq.~(\ref{eq:A_mult})) and
gives \begin{equation}
P=-2J_{\B}\sum_{\zeta=\uparrow,\downarrow}X_{\zeta}\left(e^{-2K_{\B}}\langle\zeta\rangle-\langle\bar{\zeta}\rangle\right).\label{eq:P_mult}\end{equation}
Furthermore, $P/A$ can be calculated for arbitrary dimensions and
geometries, as it is solely a property of the fluctuating field. We
find\begin{eqnarray}
\frac{P}{A} & = & -2J_{\B}\frac{\sum_{\zeta=\uparrow,\downarrow}X_{\zeta}\left(e^{-2K_{\B}}\langle\zeta\rangle-\langle\bar{\zeta}\rangle\right)}{\sum_{\zeta=\uparrow,\downarrow}X_{\zeta}\left(e^{-2K_{\B}}\langle\zeta\rangle+\langle\bar{\zeta}\rangle\right)}\nonumber \\
 & = & -J_{\B}\sum_{\zeta=\uparrow,\downarrow}\frac{e^{-2K_{\B}}\langle\zeta\rangle-\langle\bar{\zeta}\rangle}{e^{-2K_{\B}}\langle\zeta\rangle+\langle\bar{\zeta}\rangle}\nonumber \\
 & = & \frac{2J_{\B}(m^{2}+1)\tanh K_{\B}}{1-m^{2}\tanh^{2}K_{\B}}.\label{eq:P/A}\end{eqnarray}
For the magnetization Eq.~(\ref{eq:mag1d}) of the $\Se$ system
this gives \begin{equation}
\frac{P}{A}=\left\{ \begin{array}{lll}
{\displaystyle \frac{2J_{\B}e^{-4K}}{\tanh K_{\B}}} &  & T\leq\Tc\\
\\2J_{\B}\tanh K_{\B} &  & T\geq\Tc\end{array}\right.,\label{eq:P_1d}\end{equation}
which, multiplied with $A$ from Eq.~(\ref{eq:UbFit}) and for $J=J_{\B}$,
becomes\begin{equation}
P=\left\{ \begin{array}{lll}
{\displaystyle \frac{2e^{-8K}\coth2K}{\tanh K\sinh2K}} &  & T\leq\Tc\\
\\{\displaystyle \frac{2e^{-3K}\tanh K}{\cosh K}} &  & T\geq\Tc\end{array}\right..\label{eq:P_1d_1}\end{equation}
These results are shown in Fig.~\ref{fig:A_and_P_1d}, together with
data from MC simulations. Note that these results are only valid for
the multiplicative rate $p_{\mathrm{flip}}^{*}$ from Eq.~(\ref{eq:p_flip_FR}).

Finally we list the critical exponents for the $\Se$ driven system
at $v\to\infty$ to be\begin{equation}
\beta=\frac{1}{2},\;\gamma=1,\;\alpha=0.\label{eq:exponents}\end{equation}
The behavior of this system at finite velocities $v$ will be discussed
in Sec.~\ref{sec:Monte-Carlo-Simulations}.

\subsection{2d$_{\mathbf{b}}$ case}

\begin{figure}
\begin{centering}
\includegraphics[width=0.9\columnwidth]{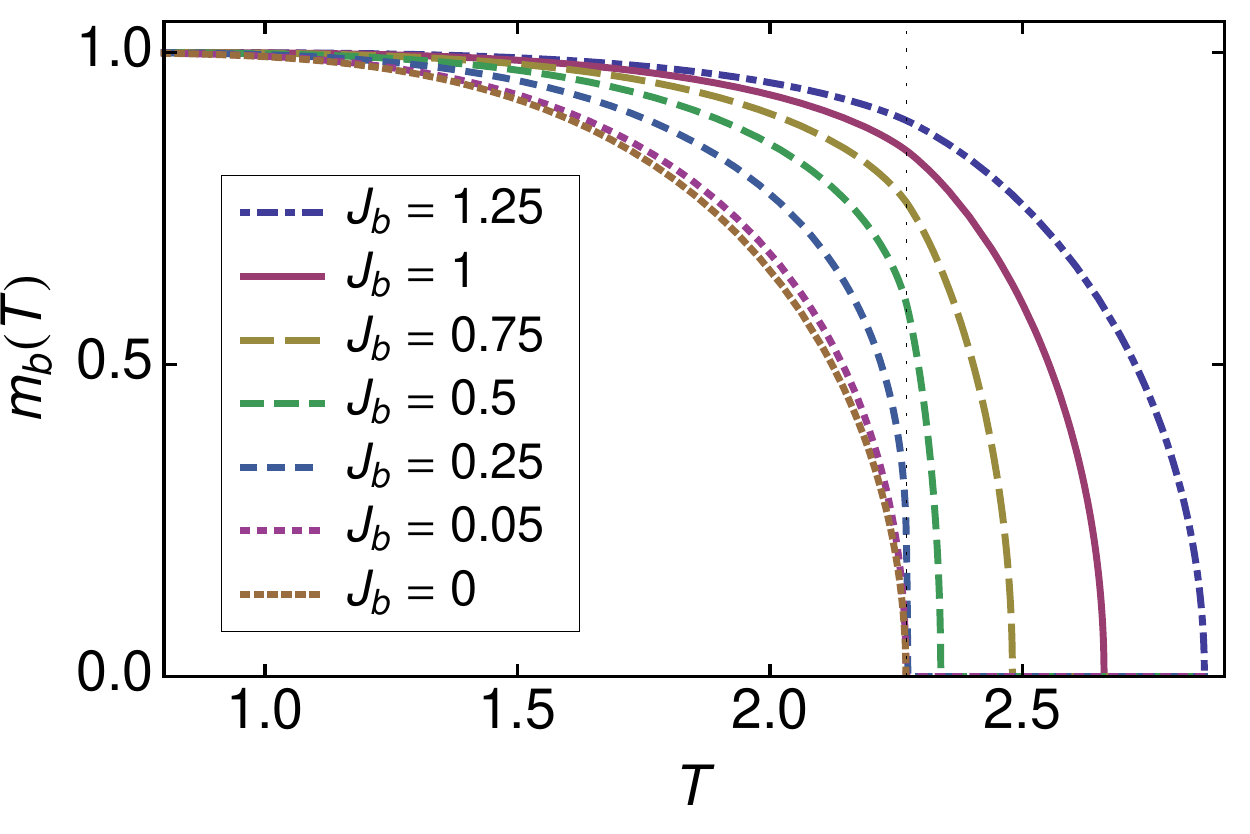}
\par\end{centering}

\caption{(Color online) Boundary magnetization $m_{\B}(T)$, Eq.~(\ref{eq:m_2d}),
of the $\Szb$ system for $J=1$ and several values of $J_{\B}$.
For $J_{\B}=0$ the $m_{\B}$ reduces to the surface magnetization
of the $2d$ equilibrium Ising model, Eq.~(\ref{eq:m_beq}). \label{fig:m_2d}}

\end{figure}
The $\Szb$ case can be solved exactly using the expression for the
equilibrium surface magnetization $m_{\B,\mathrm{eq}}(z,y_{\B})$
of the $2d$ Ising model in a static surface field $h_{\B}$ obtained
by McCoy and Wu \citep[Chapter VI, Eq. 5.1]{McCoyWu73}, with $z=\tanh K$
and $y_{\B}=\tanh h_{\B}$. The integral representation given in their
work can be further evaluated and written in closed form, the results
are given in Appendix~\ref{sec:Appendix_m_eq_2d}, Eq.~(\ref{eq:m_eq_2d}).
If we again use Eq.~(\ref{eq:sc}) and set $y_{\B}=m_{\B}z_{\B}$,
with $z_{\B}=\tanh K_{\B}$, we can calculate the non--equilibrium
boundary magnetization $m_{\B}(z,z_{\B})$ numerically as solution
of the self--consistence condition\begin{equation}
m_{\B,\mathrm{eq}}(z,m_{\B}z_{\B})=m_{\B},\label{eq:m_2d}\end{equation}
which is shown for $J=1$ and several values of $J_{\B}$ in Fig.~\ref{fig:m_2d}.
The critical temperature $\Tc$ of the system can be evaluated from
the reduced zero field boundary susceptibility $\chi_{\B,\mathrm{eq}}(z)$,
Eq. (\ref{eq:chi_eq_2d}), to give \begin{equation}
\Tc=2.6614725655752\ldots\label{eq:Tc_2db}\end{equation}
for the case $J_{\B}=J=1$ using $\chi_{\B,\mathrm{eq}}^{(0)}(z_{\mathrm{c}})z_{\B,\mathrm{c}}=1$
(Eq.~(\ref{eq:chi_Tc})). 

As the critical temperature $\Tc$, Eq.~(\ref{eq:Tc_2db}), is larger
than the equilibrium critical temperature $T_{\mathrm{c,eq}}=2.26918\ldots$,
the driven boundary induces a surface phase transition where only
the driven surface has long range order above $T_{\mathrm{c,eq}}$.
The velocity dependence of this transition and the resulting phase
diagram is discussed in more detail in Section~\ref{sec:Monte-Carlo-Simulations}.

\subsection{1+1d sheared case}

If the motion of the lattice described by Eq.~(\ref{eq:H_2d}) is
not restricted to one row but applied to the whole system we get a
system with uniform shear. Then all $\Delta_{k}(t)\equiv\Delta(t)=vt$
are equal, and we assume $K_{\perp,k}\equiv K_{\perp}$ to get \begin{equation}
\bH(t)=-\sum_{k=1}^{\Ls\vphantom{\Lp}}\sum_{l=1}^{\Lp}K_{\para}\sigma_{k,l}\sigma_{k,l+1}+K_{\perp}\sigma_{k,l}\sigma_{k+1,l+\Delta(t)}.\label{eq:H_2d_shear}\end{equation}
Note that this system is translationally invariant in both directions,
a fact that drastically simplifies the analysis of the critical behavior.

\begin{figure}
\begin{centering}
\includegraphics[scale=0.7]{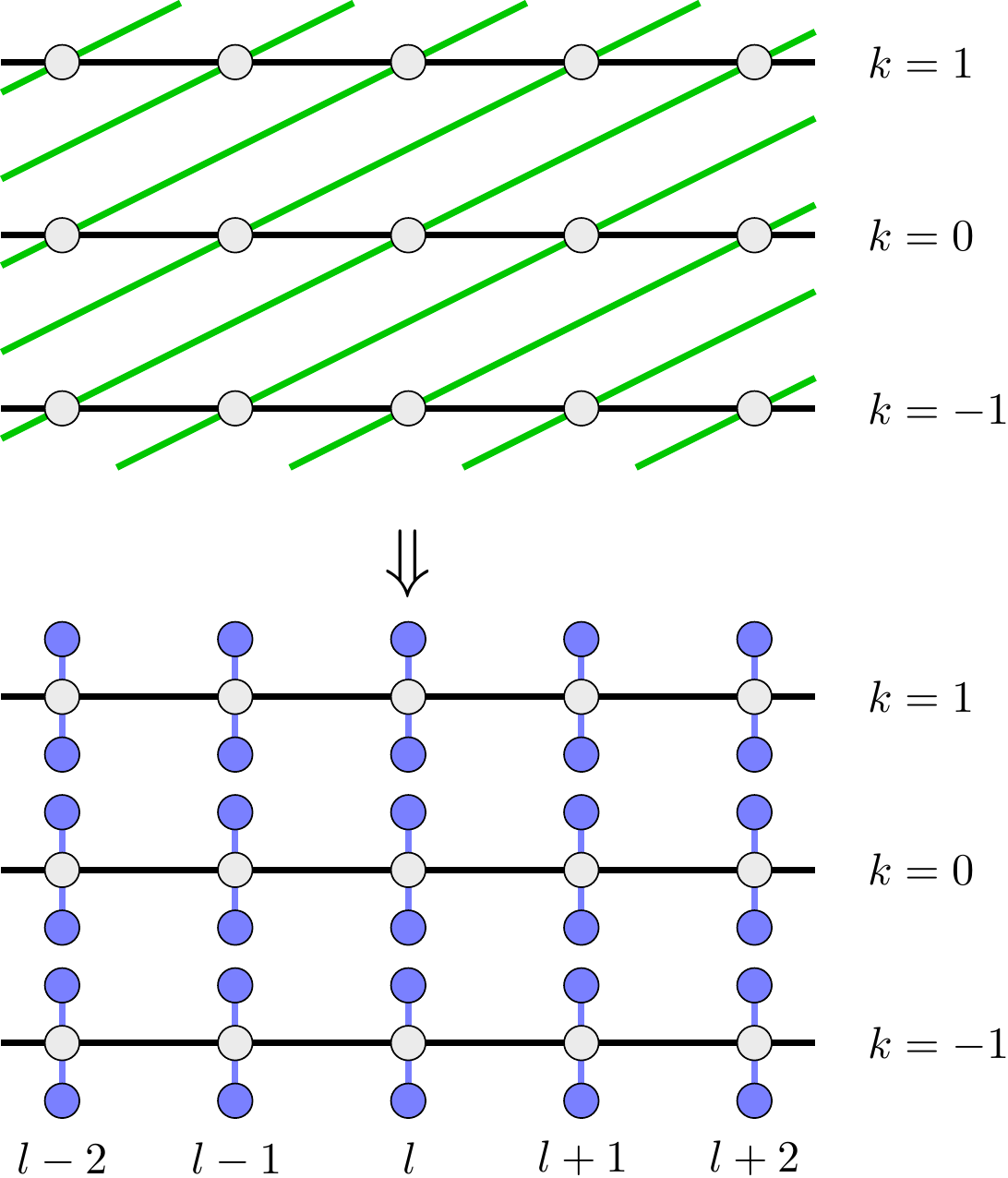}
\par\end{centering}

\caption{(Color online) Mapping of the $\See$ sheared system, shown for $\Delta=2$,
on $\Ls$ disconnected $1d$ systems with fluctuating fields \label{fig:2d_shear_mapping}}

\end{figure}
Now we will investigate this system in the limit $v\rightarrow\infty$.
Then each spin $\sigma_{kl}$ interacts, as depicted in Fig.~\ref{fig:2d_shear_mapping},
with its neighbors $\sigma_{k\pm1,l\pm\Delta(t)}$ via fluctuating
fields, while the interaction to the parallel neighbors $\sigma_{k,l\pm1}$
remains unchanged. Thus the system decomposes into $\Ls$ identical
$1d$ Ising models which again can be solved exactly: The coupling
to two fluctuating fields $\mu_{i,1}$ and $\mu_{i,2}$ with equal
strength $k_{i}$ on each site can be traced similar to Eq.~(\ref{eq:Z_tilde})
to give

\begin{eqnarray}
\mathcal{Z} & = & \Tr_{\sigma}e^{-\beta\mathcal{H}_{0}}\Tr_{\mu}\prod_{i}\prod_{j=1}^{2}e^{k_{i}\mu_{ij}\sigma_{i}}\nonumber \\
 & = & \Tr_{\sigma}e^{-\beta\mathcal{H}_{0}}\prod_{i}\prod_{j=1}^{2}\sum_{\mu_{ij}=\pm1}e^{k_{i}\mu_{ij}\sigma_{i}}p_{i}(\mu_{ij})\nonumber \\
 & = & \Tr_{\sigma}e^{-\beta\mathcal{H}_{0}}\prod_{i}\left[\cosh k_{i}+m_{i}\sigma_{i}\sinh k_{i}\right]^{2}\nonumber \\
 & = & \prod_{i}C_{i}\Tr_{\sigma}e^{-\beta\mathcal{H}_{0}}\prod_{i}\left[1+\sigma_{i}\frac{m_{i}}{C_{i}}\sinh2k_{i}\right],\label{eq:Z_tilde2}\end{eqnarray}
 with \begin{equation}
C_{i}=\frac{1}{2}\left(1-m_{i}^{2}+(1+m_{i}^{2})\cosh2k_{i}\right).\label{eq:Cmb}\end{equation}
 Equating Eq.~(\ref{eq:Z_tilde2}) with Eq.~(\ref{eq:Z_eq}) we
conclude that static fields $b_{i}$ can replace the fluctuating fields
$\mu_{ij}$, with average $m_{i}$, if \begin{equation}
\tanh b_{i}=\frac{2m_{i}\sinh2k_{i}}{1-m_{i}^{2}+(1+m_{i}^{2})\cosh2k_{i}}.\label{eq:b_condition2}\end{equation}
The sheared system is translationally invariant in both directions,
leading to homogeneous values $m_{i}=m$, $k_{i}=K_{\perp}$, and
$b_{i}=h$. Inserting Eq.~(\ref{eq:b_condition2}) into Eq.~(\ref{eq:mag1d})
we get the order parameter of the sheared $\See$ system \begin{multline}
m(K_{\para},K_{\perp})\\
=\frac{\sqrt{1-2e^{4K_{\para}}+2e^{2K_{\para}}\sqrt{e^{4K_{\para}}-1+\tanh^{2}K_{\perp}}}}{\tanh K_{\perp}}\label{eq:m_sheared}\end{multline}
with critical temperature fulfilling \begin{equation}
2e^{2K_{\para,\mathrm{c}}}\tanh K_{\perp,\mathrm{c}}=1,\label{eq:Tc_sheared}\end{equation}
which gives $\Tc=1/\log\left(\frac{1}{2}\sqrt{3+\sqrt{17}}\right)=3.46591...$
for $J_{\para}=J_{\perp}=1$. A generalization of Eq.~(\ref{eq:Z_tilde2})
from two to $f$ fluctuating fields per spin is straightforward and
leads to the general criticality condition \begin{equation}
\chi_{\mathrm{eq}}^{(0)}(K_{\mathrm{c}})\, f\tanh K_{\B,\mathrm{c}}=1.\label{eq:chi_Tc_n}\end{equation}
Although this geometry can be solved exactly at $v=\infty$ we expect
the phase transition to be \emph{strongly anisotropic} (see, e.~g.,
\citep{Hucht02a}) with two different correlation length exponents
$\nu_{\para}>\nu_{\perp}$. In fact we found such behavior, with strong
evidence for the exponents $\nu_{\para}=3/2$ and $\nu_{\perp}=1/2$,
details on this will be published elsewhere \citep{AngstHuchtWolf10}.

\subsection{Other geometries}

For two more cases we can derive highly accurate estimates for the
critical temperature $\Tc$ of the driven system when $v\rightarrow\infty$,
namely the $\Sz$ Ising double layer \citep{AngstHuchtWolf10} with
Hamiltonian \begin{multline}
\bH(t)=-\sum_{k=0}^{1\vphantom{\Lp}}\sum_{l=1}^{\Lp}\sum_{m=1}^{\Lp}\Bigl[K\sigma_{klm}(\sigma_{k,l,m+1}+\sigma_{k,l+1,m})+{}\\
{}+K_{\B}\sigma_{0lm}\sigma_{1,l+\Delta(t),m}\Bigr]\end{multline}
 and the experimentally relevant $\Sze$ sheared case\begin{multline}
\bH(t)=-\sum_{k=1}^{\Ls\vphantom{\Lp}}\sum_{l=1}^{\Lp}\sum_{m=1}^{\Lp}\Bigl[K_{\para}\sigma_{klm}(\sigma_{k,l,m+1}+\sigma_{k,l+1,m})+{}\\
{}+K_{\perp}\sigma_{klm}\sigma_{k+1,l+\Delta(t),m}\Bigr],\end{multline}
both on simple cubic lattices: With Eq.~(\ref{eq:chi_Tc_n}) we can
express $\Tc$ using the high temperature series expansion for the
reduced zero field susceptibility $\chi_{\mathrm{eq}}^{(0)}(K)$ of
the $2d$ Ising model, which was calculated to higher than $2000^{\mathrm{th}}$
order recently using a highly efficient polynomial time algorithm
\citep{Boukraa08}. Using this extremely accurate result we find,
for $J=J_{\B}=1$, the critical temperatures\begin{equation}
\Tc=4.058782423137980000987775040680\ldots\label{eq:Tc_2d}\end{equation}
for the two $\Sz$ layers, and\begin{equation}
\Tc=5.264750414514743550598017203424\ldots\label{eq:Tc_2+1d}\end{equation}
for the $\Sze$ sheared system with $f=2$ analogous to the $\See$
sheared system. Note that due to the high accuracy of the series these
values can be calculated to approximately $500$ and $700$ digits,
respectively.

Just for reference we also give the critical temperatures for two
more cases: The experimentally relevant $\Sdb$ case shown in Fig.~\ref{fig:Geometries}
as well as the quite theoretical $\Sd$ case of two three dimensional
systems in direct contact along the fourth dimension. In the $\Sdb$
case we find $\Tc=4.8(1)$ using the $8^{\mathrm{th}}$ order high
temperature series from Ref.~\citep[Tab. IV]{BinderHohenberg74},
while in the $\Sd$ case we obtain $\Tc=5.983835(1)$ using the $32^{\mathrm{th}}$
order series from \citep{ArisueFujiwara03}. 

All these higher dimensional geometries are expected to show strongly
anisotropic behavior with two ($\Sz$, $\Sdb$ and $\Sd$ case) or
possibly even three different exponents ($\Sze$ case), the reader
is referred to Ref.~\citep{AngstHuchtWolf10}.

\section{Monte Carlo Simulations\label{sec:Monte-Carlo-Simulations}}

\subsection{Method}

We now describe the algorithms used to investigate the driven system:
For finite velocities $v$ we shift the boundary couplings by increasing
$\Delta(t)$ from Eq.~(\ref{eq:shift}) after every $N/v$ random
sequential single spin flip attempts, where $N$ denotes the total
number of spins. Using $10^{5}-10^{6}$ MCS per temperature, we measured
the following boundary properties: The boundary magnetization per
spin and the energy per bond parallel to and across the boundary of
a given configuration\begin{subequations} \begin{eqnarray}
M_{\B} & = & \frac{1}{2\Lp}\sum_{k=0}^{1\vphantom{\Lp}}\sum_{l=1}^{\Lp}\sigma_{k,l}\\
E_{\B,\para} & = & -\frac{J}{2\Lp}\sum_{k=0}^{1\vphantom{\Lp}}\sum_{l=1}^{\Lp}\sigma_{k,l}\sigma_{k,l+1}\\
E_{\B} & = & -\frac{J_{\B}}{\Lp}\sum_{l=1}^{\Lp}\sigma_{0,l}\sigma_{1,l+\Delta(t)}\end{eqnarray}
\end{subequations}as well as the corresponding bulk quantities. From
these time dependent quantities we calculate the averages of the magnetization,
reduced susceptibility, Binder cumulant, internal energy and specific
heat at the boundary,\begin{subequations} \begin{eqnarray}
m_{\B,\mathrm{abs}} & = & \langle|M_{\B}|\rangle\\
\chi_{\B,\mathrm{abs}} & = & 2\Lp\left(\langle M_{\B}^{2}\rangle-\langle|M_{\B}|\rangle^{2}\right)\\
U_{\B} & = & 1-\frac{\langle M_{\B}^{4}\rangle}{3\langle M_{\B}^{2}\rangle^{2}}\label{eq:BinderUb}\\
e_{\B} & = & \langle E_{\B}\rangle\\
c_{\B} & = & \Lp\beta^{2}\left(\langle E_{\B}^{2}\rangle-\langle E_{\B}\rangle^{2}\right).\end{eqnarray}
\end{subequations} Note that we have absorbed the factor $\beta^{-1}$
into $\chi$. Near criticality these quantities show power law behavior
and fulfill \begin{subequations} \begin{eqnarray}
m_{\B,\mathrm{abs}}(\tau) & \propto & (-\tau)^{\beta}\label{eq:m_babs}\\
\chi_{\B,\mathrm{abs}}(\tau) & \propto & |\tau|^{-\gamma}\label{eq:chi_babs}\\
c_{\B}(\tau) & \propto & |\tau|^{-\alpha}\label{eq:c_b}\end{eqnarray}
\end{subequations} with reduced temperature $\tau=T/\Tc-1$ and critical
exponents $\beta$, $\gamma$ and $\alpha$. But, before we present
the results, we have to take a closer look at the used spin flip rates.

\subsection{An integrable algorithm\label{sub:An-integrable-algorithm}}

While equilibrium properties are most efficiently investigated in
Monte Carlo simulations using cluster algorithms, non--equilibrium
systems have to be treated with random sequential single spin flip
dynamics like the non-conserved Glauber dynamics \citep{Glauber63}
or the conserved Kawasaki dynamics \citep{Kawasaki65}. The driven
system is permanently under an external perturbation which drives
it out of equilibrium, while the internal degrees of freedom are coupled
to a heat bath in thermal equilibrium. From this coupling the spin
flip probability $p_{\mathrm{flip}}(\Delta E)$ of a given energy
change $\Delta E$ fulfills the detailed balance condition\begin{equation}
\frac{p_{\mathrm{flip}}(\Delta E)}{p_{\mathrm{flip}}(-\Delta E)}=e^{-\beta\Delta E}\label{eq:detailed_balance}\end{equation}
just like in the equilibrium case (for details, see \citep{AngstHuchtWolf10}).

The most common rates fulfilling Eq.~(\ref{eq:detailed_balance})
are the Metropolis rate \citep{Metropolis53} and the Glauber rate
\citep{Glauber63}, \begin{subequations}\label{eq:p_flip} \begin{eqnarray}
p_{\mathrm{flip}}^{\mathrm{M}}(\Delta E) & = & \min(1,e^{-\beta\,\Delta E}),\label{eq:p_flip_MP}\\
p_{\mathrm{flip}}^{\mathrm{G}}(\Delta E) & = & \frac{1}{1+e^{\beta\,\Delta E}}.\label{eq:p_flip_HB}\end{eqnarray}
\end{subequations} Using these rates in simulations of, e.\,g.,
the $\Se$ driven system, Eq.~(\ref{eq:H_1d}), it turns out that
for all $v>0$ the critical temperature $\Tc(v)$ depends on the used
rate (see also Fig.~\ref{fig:PhaseDiagram}): We find, for $v\to\infty$
and $J_{\B}=J=1$, the values $\Tc^{\mathrm{M}}=1.910(2)$ and $\Tc^{\mathrm{G}}=2.031(2)$
for the Metropolis and Glauber rate, respectively, while the exact
solution Eq.~(\ref{eq:Tc_1d}) of the model presented in Section~\ref{sec:Exact_Solution}
gives $\Tc=2.269...$. Note that a similar dependency was recently
found in the DLG by Kwak \emph{et al}.~\citep{KwakLandauSchmittmann04}.

\begin{figure}
\begin{centering}
\includegraphics[scale=0.5]{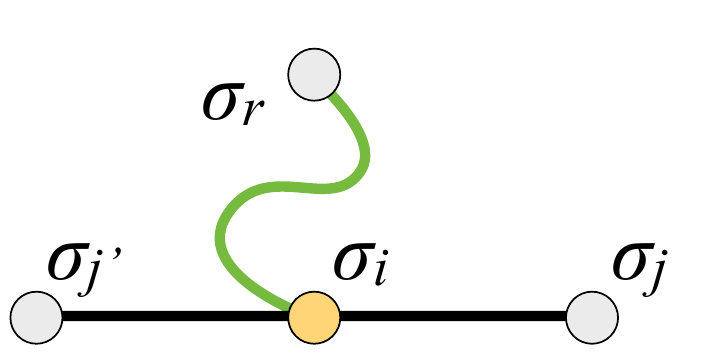}
\par\end{centering}

\caption{(Color online) Interactions of surface spin $\sigma_{i}$ in the $\Se$
case\label{fig:algorithms-picture}}

\end{figure}
\begin{figure}
\begin{centering}
\includegraphics[width=0.9\columnwidth]{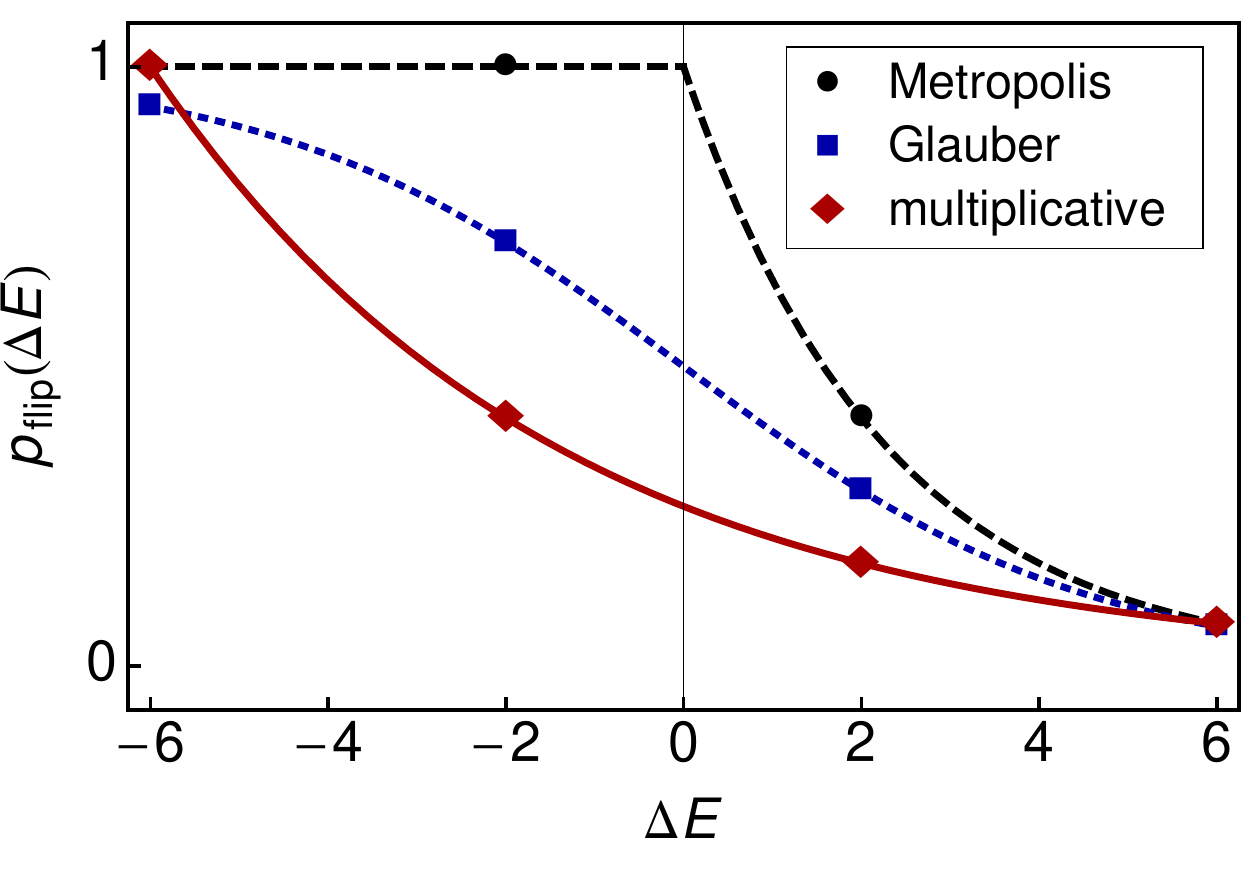}
\par\end{centering}

\caption{(Color online) Spin flip probabilities of the Metropolis rate Eq.~(\ref{eq:p_flip_MP})
(dashed black line, circles), the Glauber rate Eq.~(\ref{eq:p_flip_HB})
(dotted blue line, squares), and the multiplicative rate Eq.~(\ref{eq:p_flip_FR})
(red line, diamonds) for the $\Se$ system at criticality ($J=J_{\B}=1$).
\label{fig:algorithms}}

\end{figure}

How can these discrepancies be understood? And can we construct a
rate that matches the analytical treatment, i.\,e., has the same
$\Tc$? This is indeed possible: Consider a microscopic change, i.\,e.
a spin flip, of spin $\sigma_{i}$ at the boundary (see Fig.~\ref{fig:algorithms-picture}),
with energy difference \begin{equation}
\Delta E=\underbrace{2J\sigma_{i}\sum\nolimits _{\langle j\rangle'}^{z}\sigma_{j}}_{\Delta E_{1}}+\underbrace{2J_{\B}\sigma_{i}\vphantom{\sum\nolimits _{\langle j\rangle'}}\sigma_{r}}_{\Delta E_{2}},\label{eq:dE_example}\end{equation}
where the sum runs over the $z$ neighbors of $\sigma_{i}$ in the
same subsystem ($z=2$ in the $\Se$ case), while $\sigma_{r}$ is
from the other side of the moving boundary. The idea of the exact
solution presented in the last section was to treat spin $\sigma_{r}$
as a fluctuating variable $\mu_{i}$ at site $i$ with appropriate
statistics. By contrast, correlations of different strength are introduced
between the two subsystems by the rates Eq.~(\ref{eq:p_flip}), because
the influence of spin $\sigma_{r}$ depends on the actual state of
the $z$ spins $\sigma_{j}$. This can be seen most easily in the
case of the Metropolis rate ($J_{\B}=J$): if, e.g., $\sigma_{i}=-\sigma_{j}$
then $\Delta E_{1}=-2zJ$ and $p_{\mathrm{flip}}^{\mathrm{M}}=1$
independent of $\sigma_{r}$ (note that $\Delta E_{2}=\pm2J$), while
in the parallel case ($\sigma_{i}=\sigma_{j}$) $\Delta E_{1}=2zJ$
and $p_{\mathrm{flip}}^{\mathrm{M}}$ strongly depends on $\sigma_{r}$
(see Fig.~\ref{fig:algorithms}). 

Fortunately, these rate-induced correlations can be completely eliminated
by requiring that \emph{the flipping probability is multiplicative},
\begin{equation}
p_{\mathrm{flip}}(\Delta E_{1}+\Delta E_{2})=p_{\mathrm{flip}}(\Delta E_{1})\, p_{\mathrm{flip}}(\Delta E_{2}).\label{eq:multiplicative}\end{equation}
Clearly this condition is not satisfied for the rates in Eq.~(\ref{eq:p_flip}),
e.\,g., $ $$p_{\mathrm{flip}}^{\mathrm{M}}(-2zJ+2J)=1$, while $p_{\mathrm{flip}}^{\mathrm{M}}(-2zJ)p_{\mathrm{flip}}^{\mathrm{M}}(2J)=e^{-2K}$
(again we assume $J_{\B}=J$).

Instead, for simulations of driven systems we propose the rate \begin{equation}
p_{\mathrm{flip}}^{*}(\Delta E)=e^{-\frac{\beta}{2}(\Delta E-\Delta E_{\mathrm{min}})}\label{eq:p_flip_FR}\end{equation}
which is uniquely defined by the detailed balance condition, Eq.~(\ref{eq:detailed_balance}),
and the multiplicity condition, Eq.~(\ref{eq:multiplicative}) %
\footnote{Note that Eq.~(\ref{eq:p_flip_FR}) is mentioned in the literature
\citep{vanBeijerenSchulman84,NewmanBarkema99} without stressing the
multiplicative property, Eq.~(\ref{eq:multiplicative}).%
}. The constant $\Delta E_{\mathrm{min}}$ is the minimum possible
value of $\Delta E$ at given geometry; this assures that $p_{\mathrm{flip}}^{*}(\Delta E)$
is maximal but never larger than one. For our example Eq.~(\ref{eq:dE_example})
we find $\Delta E_{1,\mathrm{min}}=-2zJ$ and $\Delta E_{2,\mathrm{min}}=-2J_{\B}$
to fulfill Eq.~(\ref{eq:multiplicative}). This new rate reproduces
the calculated critical temperatures in all considered geometries,
e.g. $\Tc^{*}=2.269(1)$ for the $\Se$ case at $v\to\infty$. 

The resulting spin flip rates for the $\Se$ case at criticality are
shown in Fig.~\ref{fig:algorithms}. Clearly, the multiplicative
algorithm Eq.~(\ref{eq:p_flip_FR}) has a smaller overall acceptance
rate than Eqs.~(\ref{eq:p_flip}) and is thus slightly less efficient:
A finite--size scaling analysis of the acceptance rate $A=\langle p_{\mathrm{flip}}\rangle$
at criticality in the $\Se$ case yields $A_{\mathrm{c}}^{\mathrm{M}}=0.476(2)$,
$A_{\mathrm{c}}^{\mathrm{G}}=0.366(2)$ and $A_{\mathrm{c}}^{*}=0.242(2)$
for the three algorithms, rendering this method roughly two times
slower than the Metropolis algorithm. In fact, $A_{\mathrm{c}}=3\sqrt{2}-4=0.24264...$
can be calculated exactly from Eq.~(\ref{eq:A_1d1}).

Note that the Metropolis and Glauber rates can be considered as many
particle rates, as $p_{\mathrm{flip}}$ depends on the many particle
state of all coupling partners, while the multiplicative rate corresponds
to a product of two particle contributions. We believe that the dynamics
generated by the multiplicative rate is generally simpler than the
one generated by Metropolis or Glauber rates, making an exact solution
more feasible. Whether this differentiation only holds for the non-conserved
Glauber dynamics or also for the conserved Kawasaki dynamics is subject
of future work. 

In the next two sections we will investigate finite--size effects
in the $\Se$ case as well as the cross--over behavior at finite velocities
$v$ in the $\Se$ as well as in the $\Szb$ case. We first turn to
the $\Se$ case.

\subsection{1d case}

\begin{figure}
\begin{centering}
\includegraphics[width=0.9\columnwidth]{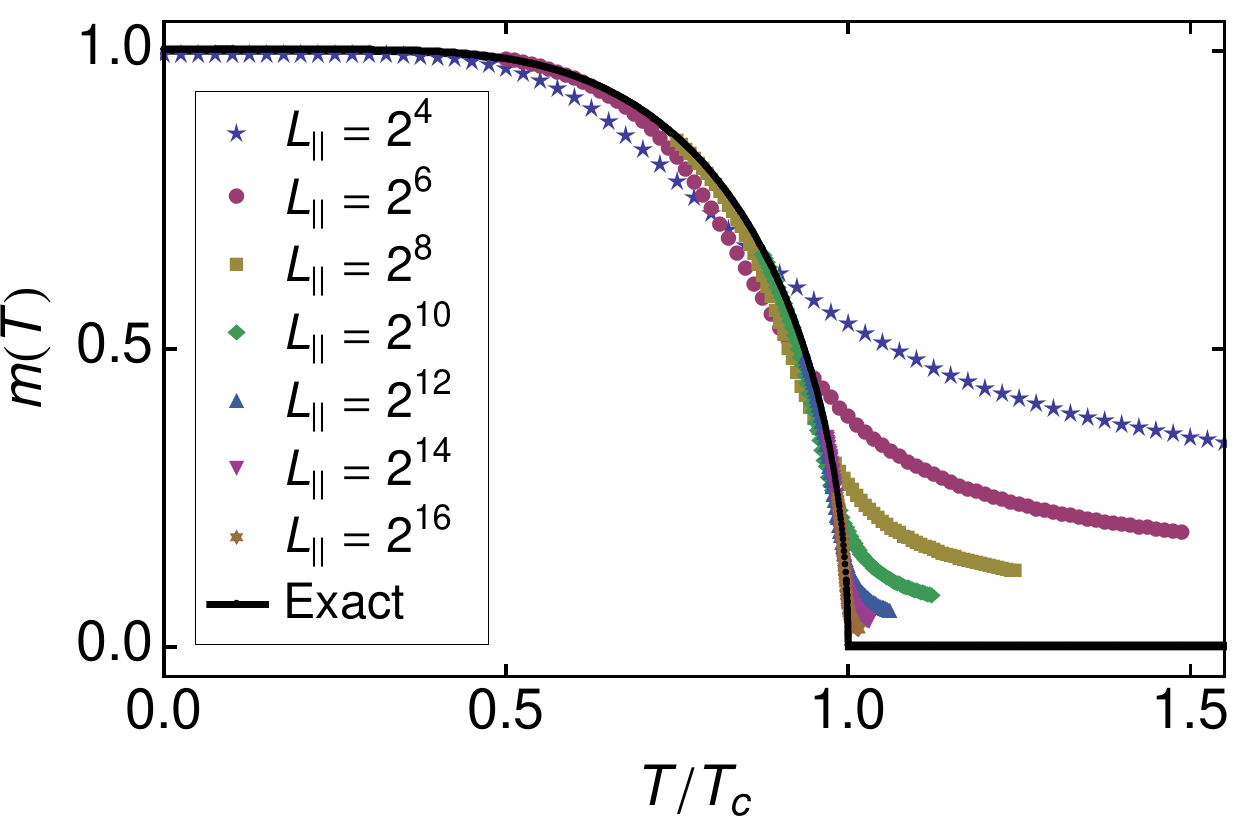}
\par\end{centering}

\caption{(Color online) Magnetization $m_{\mathrm{abs}}(T)$, Eq.~(\ref{eq:m_babs}),
of the $\Se$ system at $v=\infty$ for several system sizes $\Lp$
from Monte Carlo simulations, together with the exact solution Eq.~(\ref{eq:mag1d}).\label{fig:Magnetization1d}}

\end{figure}
The exact solution presented in Section \ref{sec:Exact_Solution}
was derived in the thermodynamic limit $\Lp\to\infty$, as we assumed
a constant and non-fluctuating order parameter $m$ in the self--consistence
condition Eq.~(\ref{eq:sc}). This led to the result that the correlation
length $\xi_{\para}$, Eq.~(\ref{eq:xi}), remains finite at criticality.
However, in a finite system the assumption $m=\mathit{const}$ is
not fulfilled and finite--size effects occur, leading to a non trivial
dependency of the physical quantities on system size. The fluctuating
order parameter gives rise to additional correlations between spins
at large distances not included in the exact solution. As the driven
system shows mean field behavior, we can use the standard finite--size
scaling theory for mean field systems: Near criticality the correlation
length parallel to the boundary fulfills $\xi_{\para}(\tau)\propto|\tau|^{-\nu_{\para}}$
with critical exponent $\nu_{\para}=2/d_{\B},$ where $d_{\B}$ denotes
the boundary dimension. We have $d_{\mathrm{b}}=1$ in both the $\Se$
and the $\Szb$ case, leading to $\nu_{\para}=2$ in these cases.

To illustrate these finite--size effects in the $\Se$ case, in Fig.~\ref{fig:Magnetization1d}
we show the magnetization $m_{\mathrm{abs}}(T)$, Eq.~(\ref{eq:m_babs}),
as function of temperature for $v=\infty$ and several system sizes
$\Lp$. The exact solution, Eq.~(\ref{eq:mag1d}), is only approached
in the limit $\Lp\to\infty$.

\begin{figure}
\begin{centering}
\includegraphics[width=0.9\columnwidth]{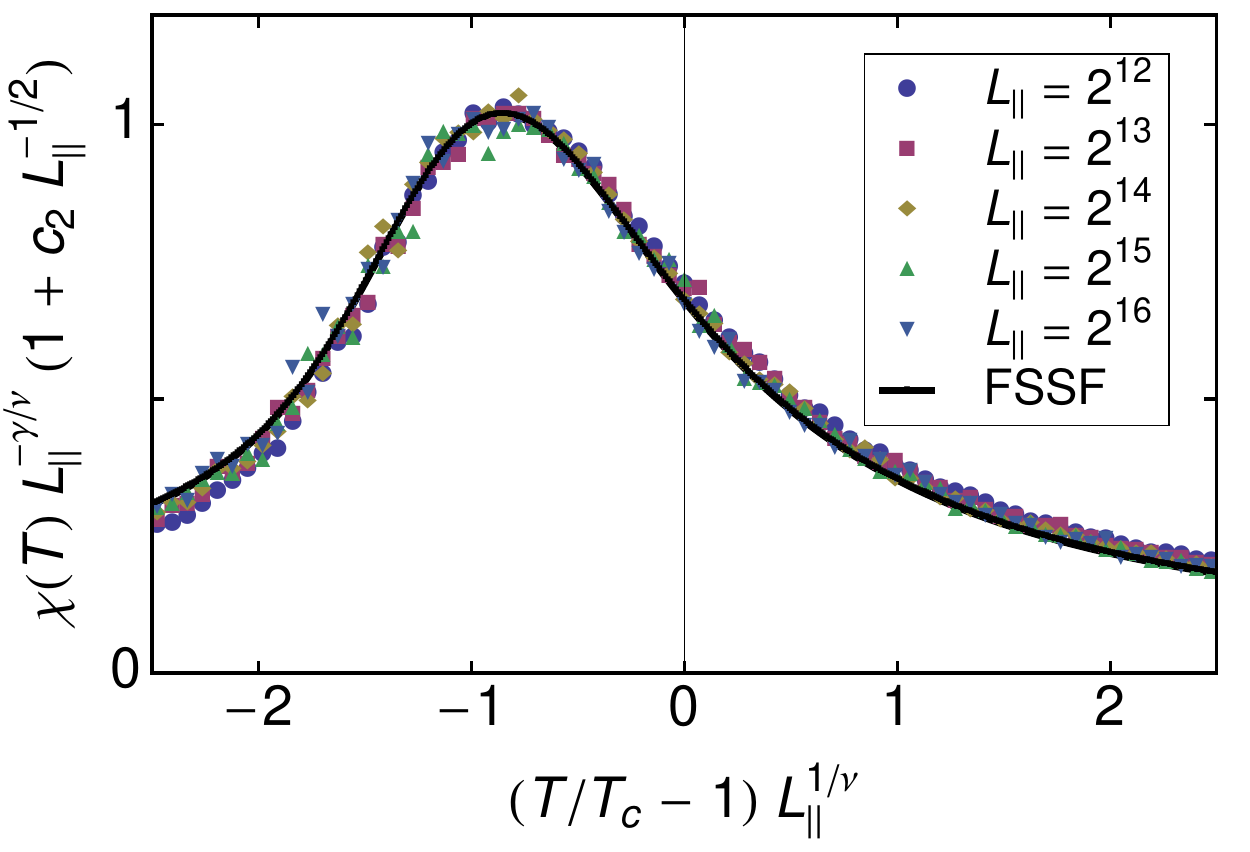}
\par\end{centering}

\caption{(Color online) Finite--size scaling plot of the reduced susceptibility
$\chi_{\mathrm{abs}}(T)$, Eq.~(\ref{eq:chi_babs}), of the $\Se$
system for $v=\infty$ and system sizes $\Lp=2^{12},\ldots,2^{16}$,
together with the exact mean field finite--size scaling function (black
line) from Ref.~\citep{GruenebergHucht04}. The correction factor
$c_{2}=2.7$. \label{fig:Chi1d}}

\end{figure}
The finite--size scaling behavior is demonstrated exemplarily for
the susceptibility $\chi_{\mathrm{abs}}(T)$, Eq.~(\ref{eq:chi_babs}),
which is shown in a finite--size scaling plot in Fig.~\ref{fig:Chi1d}:
After rescalation of the MC data in the usual way we indeed find the
expected mean field exponents $\gamma=1$ and $\nu_{\para}=2$, furthermore
the data falls onto the universal finite--size scaling function calculated
in Ref.~\citep{GruenebergHucht04}. The same analysis was performed
for the magnetization $m_{\mathrm{abs}}(T)$ and specific heat $c(T)$,
Eq.~(\ref{eq:c_b}), verifying the other two exponents $\beta=1/2$
and $\alpha=0$. 

In summary, the $\Se$ and the $\Szb$ systems with boundary dimension
$d_{\mathrm{b}}=1$ have the standard mean field exponents and fulfill
the exponent relations \begin{equation}
2-\alpha=2\beta+\gamma=d_{\mathrm{b}}\nu_{\para}.\label{eq:scaling_relations}\end{equation}

\begin{figure}
\begin{centering}
\includegraphics[width=8cm]{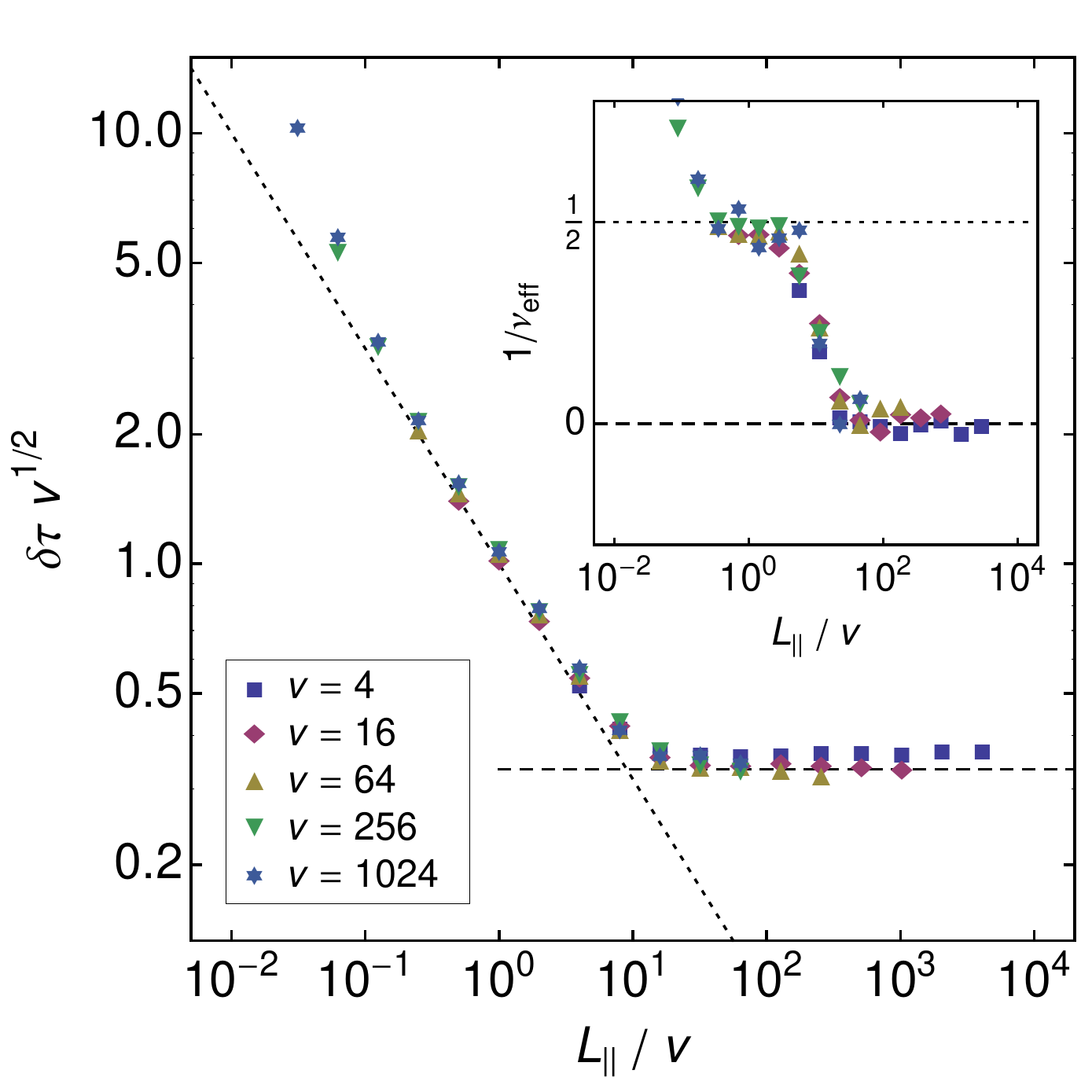}
\par\end{centering}

\caption{(Color online) Velocity dependent cross--over behavior in the $\Se$
case. Shown is the rescaled width of the critical region $\delta\tau\, v^{1/2}$
as function of the cross--over scaling variable $\Lp/v$ for several
velocities $v$ and several system sizes $\Lp=2^{4},\ldots,2^{16}$
(see text). The inset shows the corresponding cross--over of the effective
correlation length exponent $\nu_{\mathrm{eff}}^{-1}$ from $\nu_{\mathrm{eff}}^{-1}=1/2$
(MF, dotted line) to $\nu_{\mathrm{eff}}^{-1}=0$ (Ising non-critical,
dashed line).\label{fig:CrossOver1d}}

\end{figure}
We now turn to finite velocities $v$: Then the $\Se$ system always
shows a cross--over from mean field to Ising behavior with increasing
system size $\Lp$. Only in the limit $v\rightarrow\infty$ the system
undergoes a phase transition at finite temperatures. To investigate
this velocity dependent cross--over, we measured the width $\delta\tau$
of the critical region by analysing the Binder cumulant Eq.~(\ref{eq:BinderUb}).
Using least square fits of the simulation data to the simple approximation
\begin{equation}
U_{\B}(T)\approx\left\{ \begin{array}{lll}
{\displaystyle \frac{1}{3}[1+\tanh(\tilde{\tau}/\delta\tau)]} &  & \tilde{\tau}\leq0\\
\\{\displaystyle \frac{1}{3}\frac{1}{1+\tilde{\tau}/\delta\tau}} &  & \tilde{\tau}>0\end{array}\right.,\label{eq:UbFit}\end{equation}
with $\tilde{\tau}=T/\tilde{T}_{\mathrm{c}}-1$ and fit parameters
$\tilde{T}_{\mathrm{c}}$ and $\delta\tau$, for several velocities
$v$ and system sizes $\Lp$ we determined $\delta\tau$ and plotted
them in Fig.~\ref{fig:CrossOver1d}. We find that the cross--over
scaling variable is $\Lp/v$ in this case, while the $y$-axis has
to be rescaled as $\delta\tau\, v^{1/2}$ to get the correct limit
$\Lp^{1/\nu_{\para}}\,\delta\tau=\mathit{const}$ with $\nu_{\para}=2$
in the limit $v\to\infty$. At finite $v$ the width $\delta\tau$
stops shrinking at $L_{\para}^{\times}\approx9v$, where $L_{\para}^{\times}$
denotes the cross--over system size, and only goes to zero for $v\to\infty$,
indicating a sharp phase transition in this limit. The inset shows
the effective exponent $\nu_{\mathrm{eff}}$ obtained from the logarithmic
derivative, \begin{equation}
\nu_{\mathrm{eff}}^{-1}=-\frac{\partial\log\delta\tau}{\partial\log\Lp},\label{eq:nu_eff}\end{equation}
 whose value changes from $\nu_{\mathrm{eff}}^{-1}=1/2$ (MF) to $\nu_{\mathrm{eff}}^{-1}=0$
(Ising) with growing system size. In the next section we will see
that this behavior changes substantially in the $\Szb$ case.

\subsection{2d$_{\mathbf{b}}$ case}

In the $\Szb$ case the moving boundary is coupled to a two-dimensional
Ising model, which undergoes a phase transition at $T_{\mathrm{c,eq}}$,
Eq.~(\ref{eq:Tc_Ising}), independent of the velocity $v$. In addition,
the moving boundary shows a boundary phase transition at temperature
$\Tc(v)$, which grows with increasing $v$ and eventually approaches
the value given in Eq.~(\ref{eq:Tc_2db}) for $v\to\infty$. As $\Tc(v)>T_{\mathrm{c,eq}}$
for all $v>0$ we expect a boundary phase transition with paramagnetic
bulk. Then the correlation length $\xi_{\perp}$ perpendicular to
the boundary is finite at criticality and has the Ising value \begin{equation}
\xi_{\perp,\mathrm{c}}(v)=\xi_{\mathrm{eq}}(\Tc(v)),\label{eq:xi_perpc}\end{equation}
 with \citep{McCoyWu73} \begin{equation}
\xi_{\mathrm{eq}}^{-1}(T)=\left\{ \begin{array}{ll}
4K-2\log\coth K & \qquad T<T_{\mathrm{c,eq}}\\
\log\coth K-2K & \qquad T>T_{\mathrm{c,eq}}\end{array}\right..\label{eq:xi_2d_eq}\end{equation}
For that reason, in the finite--size scaling analysis it is sufficient
for given $v$ to simulate systems with varying length $\Lp$ while
holding the height $\Ls$ fixed at a value $\Ls\gg\xi_{\perp,\mathrm{c}}(v)$.
\begin{figure}
\begin{centering}
\includegraphics[width=0.9\columnwidth]{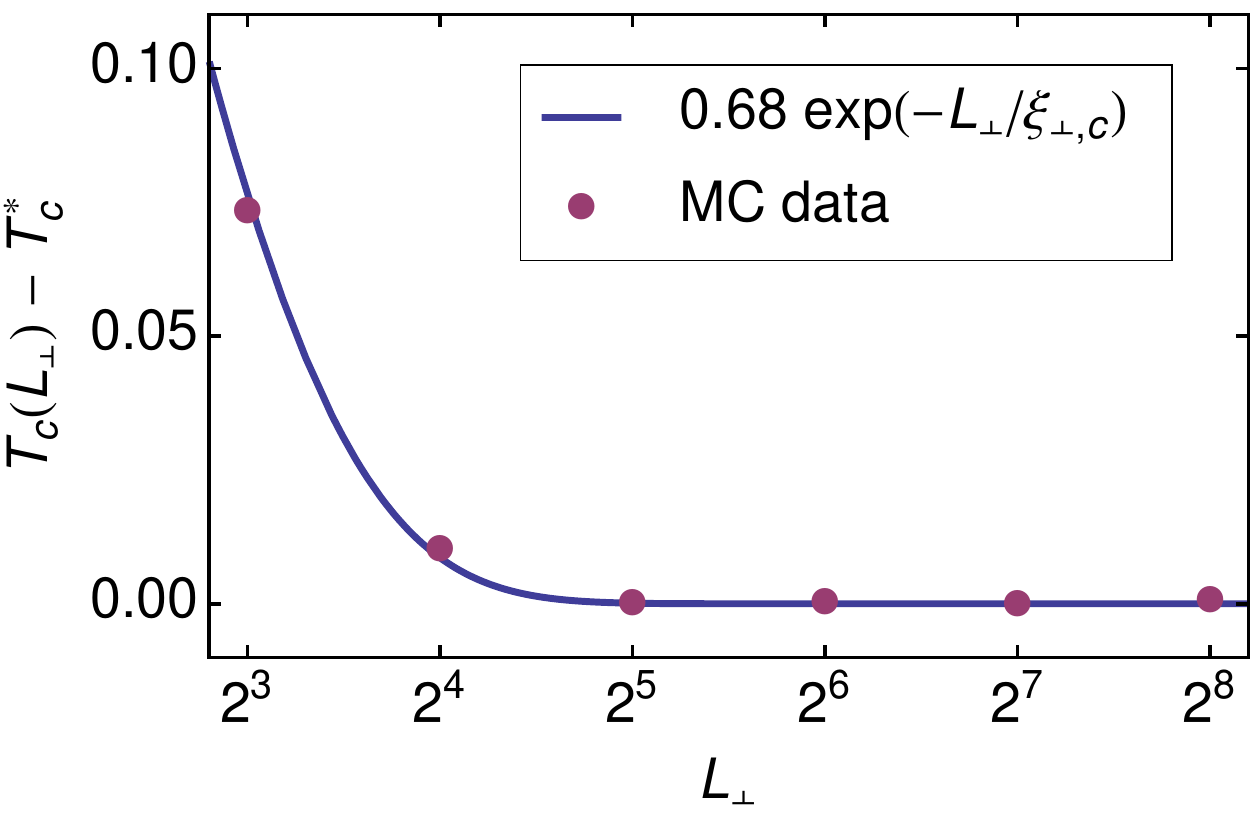}
\par\end{centering}

\caption{(Color online) Influence of the system size $\Ls$ on the critical
point in the $\Szb$ case at $v=\infty$ and $\Lp=256$. The effective
critical temperature $\Tc(\Ls)$ shifts to higher values if $\Ls\lesssim10\xi_{\perp,\mathrm{c}}$
(see text). \label{fig:TcShift}}

\end{figure}
To illustrate this behavior, in Fig.~\ref{fig:TcShift} we show the
effect of different values of $\Ls$ for $v=\infty$ and $\Lp=256$.
Only below $\Ls\approx32$ the system feels the finite width $\Ls$,
resulting in a shift of the effective critical temperature $\Tc(\Ls)$
to higher values. The strength of the shift is proportional to the
correlation function in $\perp$ direction, $\langle\sigma_{0,l}\sigma_{\Ls,l}\rangle\propto\exp(-\Ls/\xi_{\perp,\mathrm{c}})$.
The curves collapse for $\Ls>32$ showing that a ratio $\Ls/\xi_{\perp,\mathrm{c}}\approx10$
is sufficient, as $\xi_{\perp,\mathrm{c}}(\infty)=3.66323\ldots$
in this case. 

\begin{table}[b]
\begin{centering}
\begin{tabular}{>{\centering}p{0.2\columnwidth}>{\centering}p{0.2\columnwidth}>{\centering}p{0.2\columnwidth}}
$v$ & $\Tc^{*}(v)$ & $\Tc^{\mathrm{M}}(v)$\tabularnewline
\hline
$1/16$ & $2.301(2)$ & \tabularnewline
$1/4$ & $2.33(1)\hphantom{0}$ & \tabularnewline
$1$ & $2.41(1)\hphantom{0}$ & $2.30(2)$\tabularnewline
$4$ & $2.52(1)\hphantom{0}$ & $2.37(2)$\tabularnewline
$16$ & $2.61(1)\hphantom{0}$ & $2.42(2)$\tabularnewline
$64$ & $2.644(3)$ & $2.44(2)$\tabularnewline
$256$ & $2.654(2)$ & $2.44(2)$\tabularnewline
$1024$ & $2.659(2)$ & $2.45(2)$\tabularnewline
$\infty$ & $2.661(1)$ & $2.45(2)$\tabularnewline
\end{tabular}
\par\end{centering}

\caption{Velocity dependent critical temperatures $\Tc(v)$ for the $\Szb$
case, estimated using the multiplicative rate, Eq.~(\ref{eq:p_flip_FR}),
with 400.000 MC sweeps per temperature as well as using the Metropolis
rate, Eq.~(\ref{eq:p_flip_MP}), with 50.000 MC sweeps per temperature.
\label{tab:Tcs}}

\end{table}
\begin{figure}
\begin{centering}
\includegraphics[width=0.9\columnwidth]{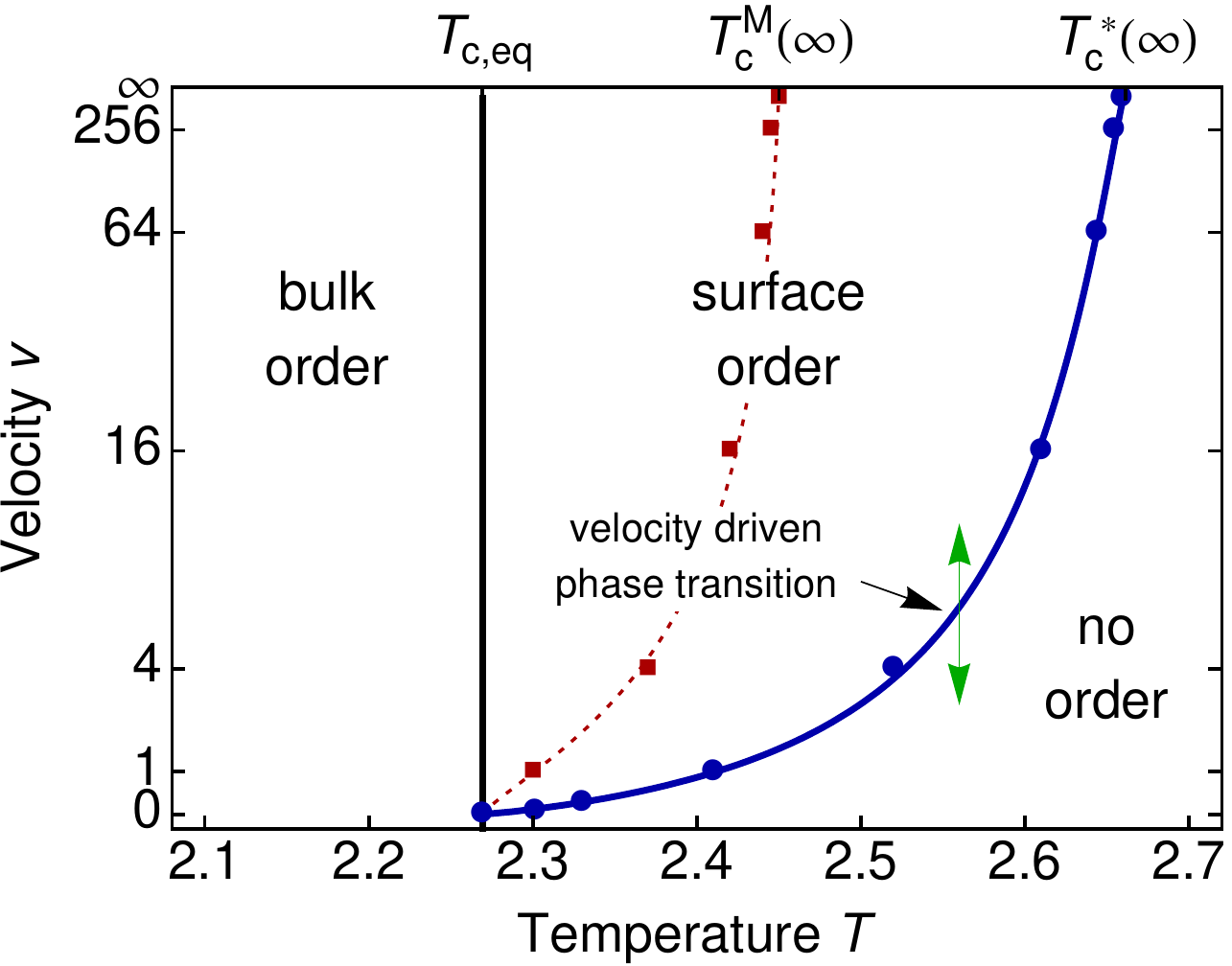}
\par\end{centering}

\caption{(Color online) Phase diagram of the $\Szb$ case. Below $T_{\mathrm{c,eq}}$
the two-dimensional bulk is ordered, while surface order is observed
even above $T_{\mathrm{c,eq}}$ up to the velocity dependent phase
boundary $\Tc(v)$. The position of this boundary depends on the algorithm,
the blue line holds for the multiplicative rate, Eq.~(\ref{eq:p_flip_FR}),
while the thin red dotted line holds for the Metropolis rate, Eq.~(\ref{eq:p_flip_MP}).
At fixed temperatures between $T_{\mathrm{c,eq}}$ and $\Tc(v)$ a
velocity driven phase transition is possible. The points are results
from MC simulations.\label{fig:PhaseDiagram}}

\end{figure}
We performed MC simulations and determined the critical temperatures
for different velocities $v$ by performing a finite--size scaling
analysis of the boundary susceptibility $\chi_{\mathrm{abs,b}}(T)$
and the boundary cumulant $U_{\mathrm{b}}(T)$. For the multiplicative
algorithm, Eq.~(\ref{eq:p_flip_FR}), we used 400.000 MC steps per
temperature, while for the Metropolis algorithm 50.000 MC steps per
temperature were used. The results are given in Tab.~\ref{tab:Tcs}
and are compiled into a phase diagram of the $\Szb$ case shown in
Fig.~\ref{fig:PhaseDiagram}. An important aspect of this phase diagram
is the possibility of a velocity driven non--equilibrium phase transition
at fixed temperature (double arrow): While the system is paramagnetic
at $v=0$ and up to $v_{\mathrm{c}}(T)$ (thick blue line), the boundary
shows long range order above that velocity. It would be interesting
to see this transition in experiments, which could be performed in
the corresponding geometry $\Sdb$ (see Fig.~\ref{fig:Geometries}),
e.g., using two close rotating magnets slightly above the Curie temperature.
The magnets should be isolating to avoid Eddy currents \citep{KadauHuchtWolf08}.

\begin{figure}
\begin{centering}
\includegraphics[width=8cm]{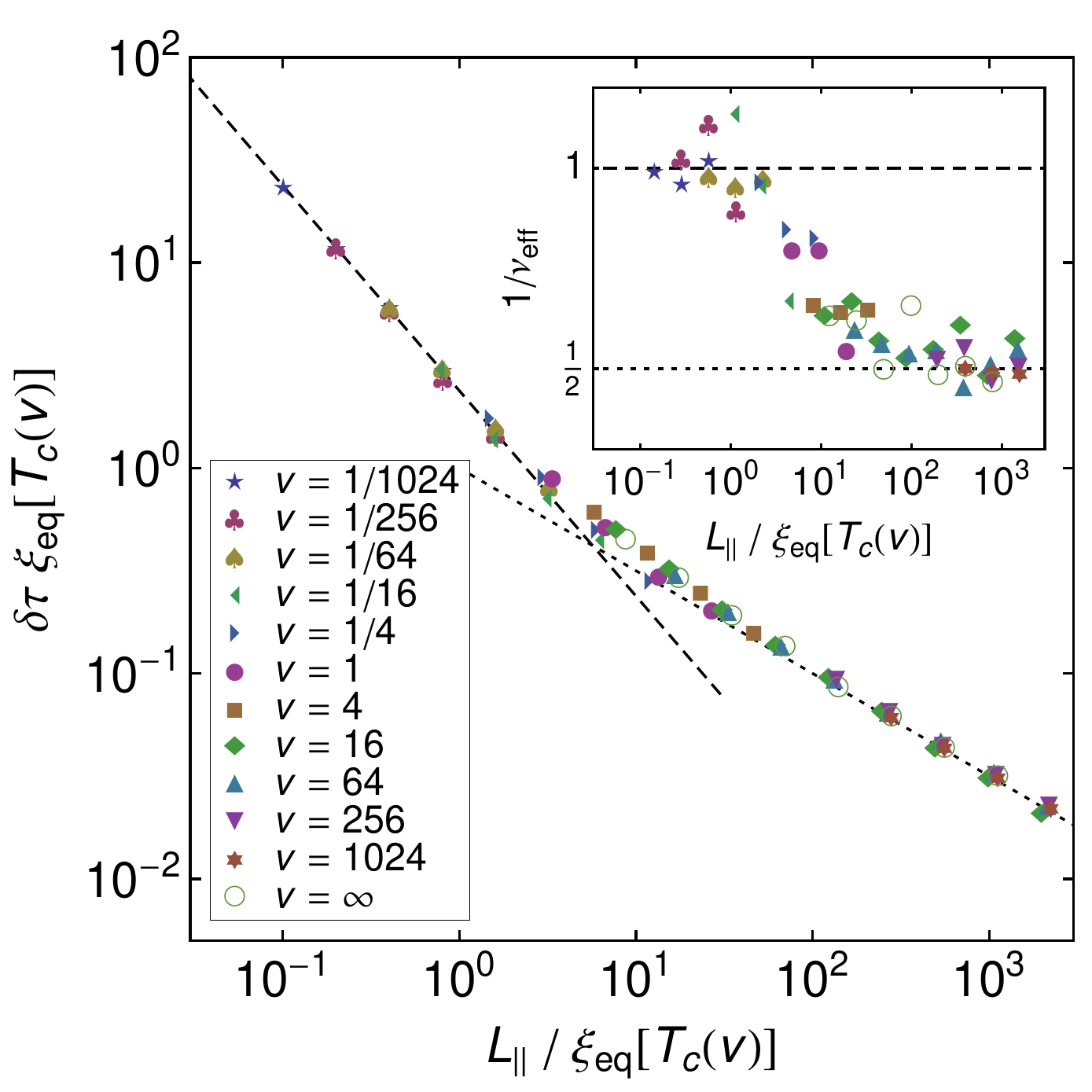}
\par\end{centering}

\caption{(Color online) Velocity dependent cross--over behavior in the $\Szb$
case. Shown is the rescaled width of the critical region $\delta\tau\,\xi_{\mathrm{eq}}[\Tc(v)]$
as function of the cross--over scaling variable $\Lp/\xi_{\mathrm{eq}}[\Tc(v)]$
for several velocities $v$ and different system sizes $\Lp=2^{4},\ldots,2^{10}$
(see text). The inset shows the corresponding cross--over of the effective
correlation length exponent $\nu_{\mathrm{eff}}$ from $\nu_{\mathrm{eff}}=1$
(Ising, dashed line) to $\nu_{\mathrm{eff}}=2$ (MF, dotted line).\label{fig:CrossOver2db}}

\end{figure}
In the $\Szb$ case the cross--over scaling variable can be determined
from the $\Tc(v)$ dependency discussed above. The correlation length
$\xi_{\mathrm{eq}}$ at the critical point of the driven system, $\Tc(v)$,
plays a key role: The system is Ising-like as long as correlations
span the whole system in both directions $\para$ and $\perp$, i.e.
as long as the system size $\Lp$ is of the order of the bulk correlation
length $\xi_{\mathrm{eq}}$ at the critical point $\Tc(v)$ of the
driven system, leading to the cross--over scaling variable $\Lp/\xi_{\mathrm{eq}}[\Tc(v)].$
Again, the rescaling of the $y$-axis can be determined by requiring
that a data collapse is obtained in the limit $v\to0$, leading to
the expression $\delta\tau\,\xi_{\mathrm{eq}}[\Tc(v)]$, as $\xi_{\mathrm{eq}}$
cancels in this case and we get the required condition $\Lp\,\delta\tau=\mathit{const}$,
as $\xi_{\mathrm{eq}}\propto\tau^{-\nu_{\mathrm{eq}}}$ in this limit,
and $\nu_{\mathrm{eq}}=1$. 

The resulting cross--over scaling plot is shown in Fig.~\ref{fig:CrossOver2db}.
For all finite $v>0$ the critical behavior changes from Ising to
mean field at the cross--over system size $L_{\para}^{\times}\approx6\xi_{\mathrm{eq}}[\Tc(v)]$:
Below this value $\delta\tau$ shrinks according to $\delta\tau\propto\Lp^{-1}$
(Ising, dashed line), while above this value $\delta\tau\propto\Lp^{-1/2}$
holds (MF, dotted line). As the shift exponent $\theta$ at small
velocities, defined by\begin{equation}
\Tc(v)-\Tc(0)\propto v^{\theta},\label{eq:shift-exponent}\end{equation}
 is close to $1/2$ we have, for small $v$, $L_{\para}^{\times}\propto v^{-\theta}\approx v^{-1/2}$.
The shift exponent $\theta=1/2$ has also been found in a field theoretical
calculation of the $\Sze$ system \citep{GonnellaPellicoro00}.

\section{Summary}

In this work we investigated a recently proposed driven Ising model
with friction due to magnetic correlations. The non--equilibrium phase
transition present in this system was investigated in detail using
analytical methods and Monte Carlo simulations. In the far from equilibrium
limit of high driving velocities $v\to\infty$ the model was solved
exactly by integrating out the non--equilibrium degrees of freedom.
The resulting exact self--consistence equation was analysed for various
geometries, leading in many cases to precise values of the critical
temperature $\Tc$ of the non--equilibrium phase transition. In the
limit $v\to\infty$ the system always shows mean field behavior due
to dimensional reduction, independent of geometry. In the simplest
one dimensional case denoted $\Se$ a complete analysis of both equilibrium
as well as non--equilibrium quantities has been presented. These exact
results are another example of mean field critical behavior in an
exactly solvable driven system, just as in the case of the DLG in
a certain limit \citep{vanBeijerenSchulman84}.

The analytic results were reproduced using a multiplicative Monte
Carlo rate originally introduced in \citep{vanBeijerenSchulman84},
which eliminates correlations due to many particle dynamics introduced
by the common Metropolis and Glauber rates. We claim that this algorithm
is generally favorable to the Metropolis and Glauber rates if an analytical
treatment is considered. 

The finite--size effects naturally emerging in the simulations were
analyzed using finite--size scaling techniques, a perfect agreement
with exactly known universal finite--size scaling functions \citep{GruenebergHucht04}
were found. 

We analysed the critical behavior at finite velocities and studied
the cross--over behavior from low to high velocities: We found that
the $\Se$ system only has a phase transition in the thermodynamic
limit for $v=\infty$, while systems with finite $v$ always become
non--critical at the cross--over system size $\Lp^{\times}\approx9v$.
On the contrary, the two--dimensional case $\Szb$ already has an
Ising type phase transition at $v=0$, which changes to mean field
behavior for \emph{all} finite $v>0$ in the thermodynamic limit,
at a cross--over length $L_{\para}^{\times}\approx6\xi_{\mathrm{eq}}[\Tc(v)]$.
In this sense, the velocity $v$ is a relevant perturbation, always
driving the system to a non--equilibrium state.

The $\Se$ system changes from mean field to non--critical Ising universality,
while the $\Szb$ case changes from Ising to mean field type with
growing system size $\Lp$. This somewhat puzzling fact can be understood
in terms of the critical width $\delta\tau$ of the transition as
follows: As in general $\delta\tau\propto\Lp^{-1/\nu}$ at criticality,
in the two dimensional Ising case $\delta\tau\propto\Lp^{-1}$, while
in the mean field case with one dimensional boundary $\delta\tau\propto\Lp^{-1/2}$.
Thirdly, $\delta\tau\propto\Lp^{0}=\mathit{const}$ in the $\Se$
case at finite $v$. In the cross--over the actual critical width
$\delta\tau$ is always governed by the largest contribution, and
so at sufficiently large system size $\Lp$ the contribution with
smallest $\nu^{-1}$ dominates and determines the critical behavior.
As consequence in both cases the effective inverse correlation length
exponent $\nu_{\mathrm{eff}}^{-1}$ changes from a larger value at
small $\Lp$ to a smaller value at large $\Lp$ ($1/2\to0$ in the
$\Se$ case, $1\to1/2$ in the $\Szb$ case).

Comparing the results to the\emph{ }driven lattice gas (DLG) \citep{KatzLebowitzSpohn83},
we note that the DLG also shows a continuous non--equilibrium phase
transition from an ordered to a disordered state at a critical temperature
which grows with growing driving field. However, in the DLG the particle
number is conserved, while we deal with a non-conserved magnetization. 

Finally some remarks on strongly anisotropic critical behavior: The
sheared system denoted $\See$ shows strongly anisotropic behavior
at criticality and $v\to\infty$, with strong evidence for the correlation
length exponents $\nu_{\para}=3/2$ and $\nu_{\perp}=1/2$, details
on this will be published elsewhere \citep{AngstHuchtWolf10}. Remarkably,
this is a rare case of an exactly solvable non-equilibrium system
with strongly anisotropic critical behavior. 
\begin{acknowledgments}
Special thanks go to Dietrich E. Wolf for very valuable discussions,
criticism and comments within the framework of the Sonderforschungsbereich
616, \textquotedblleft{}Energy Dissipation at Surfaces\textquotedblright{}.
Thanks also to Sebastian Angst, Lothar Brendel and Felix Schmidt for
helpful discussions and to Sven Lübeck for critical reading of the
manuscript.
\end{acknowledgments}
\appendix

\section{Surface magnetization of the $2d$ Ising model \label{sec:Appendix_m_eq_2d}}

The equilibrium surface magnetization $m_{\B,\mathrm{eq}}$ of the
$2d$ Ising model in a static surface field $h_{\B}$ obtained by
McCoy and Wu \citep[Chapter VI, Eq. 5.1]{McCoyWu73} as well as the
reduced zero field boundary susceptibility \begin{equation}
\chi_{\B,\mathrm{eq}}^{(0)}=\left.\frac{\partial m_{\B,\mathrm{eq}}}{\partial h_{\mathrm{b}}}\right|_{h_{\mathrm{b}}\rightarrow0}\end{equation}
 can be written in closed form not present in the literature yet \citep{McCoy08-privcomm}.
$\chi_{\B,\mathrm{eq}}^{(0)}$ is sometimes denoted $\chi_{11}$,
and a high temperature series expansion was derived up to $10^{\mathrm{th}}$
order in Ref.~\citep{BinderHohenberg72} and up to $23^{\mathrm{th}}$
order in Ref.~\citep{EntingGuttmann80}. As the expressions for anisotropic
couplings $K_{\para}$ and $K_{\perp}$ become way too complicated,
we only give the results for the isotropic Ising model with $K_{\para}=K_{\perp}=K$
here: Using the definitions $z=\tanh K$, $y=\tanh h_{\B}$ we find
\begin{widetext}\begin{eqnarray}
m_{\B,\mathrm{eq}}(z,y) & = & \frac{z^{-1}-z}{\frac{z}{y}-\frac{y}{z}}\left[\frac{b^{2}}{2\pi}\EpK(16w^{2})+\frac{b^{2}}{4\pi w}\frac{\bigl(a+\frac{y^{2}}{z}\bigr)^{2}}{1-\frac{by^{2}}{c^{2}z}}\EpPi\!\left(\frac{\bigl(1-\frac{by^{2}}{z}\bigr)^{2}}{1-\frac{by^{2}}{c^{2}z}},16w^{2}\right)+\frac{Y^{1/2}-Y^{-1/2}}{2(z^{-1}-z)}-\frac{1}{4}\right],\label{eq:m_eq_2d}\\
\nonumber \\\chi_{\B,\mathrm{eq}}^{(0)}(z) & = & \left(\frac{1}{z^{2}}-1\right)\left[\left(1+2w-8w^{2}\right)\frac{\EpK(16w^{2})}{4\pi w}-\frac{\EpE(16w^{2})}{4\pi w}-\frac{1}{4}\right],\label{eq:chi_eq_2d}\end{eqnarray}
\end{widetext} with the abbreviations\begin{subequations}\begin{eqnarray}
w & = & \frac{z(1-z^{2})}{(1+z^{2})^{2}}\label{eq:w}\\
a & = & \frac{1-2z-z^{2}}{1+z^{2}}\\
b & = & \frac{1+2z-z^{2}}{1+z^{2}}\\
c & = & \frac{2z}{1+z^{2}}\\
Y & = & \left(\frac{az}{c^{2}y^{2}}+1\right)\left(\frac{by^{2}}{c^{2}z}-1\right)^{-1}\end{eqnarray}
\end{subequations} and the complete elliptic integrals %
\footnote{We use the definition of elliptic functions without square, e.g.,
$\EpK(m)=\int_{0}^{\pi/2}(1-m\sin^{2}\theta)^{-1/2}\mathrm{d}\theta$%
} of the $1^{\mathrm{st}}$, $2^{\mathrm{nd}}$ and $3^{\mathrm{rd}}$
kind, $\EpK(m)$, $\EpE(m)$ and $\EpPi(n,m)$ Note that the variable
$w$ is also used in high temperature series analysis of the bulk
zero field susceptibility \citep{Boukraa08}. For $h_{\B}=0$ the
surface magnetization Eq.~(\ref{eq:m_eq_2d}) reduces to the well
known expression \begin{equation}
m_{\B,\mathrm{eq}}(K)=\sqrt{\frac{\cosh2K-\coth2K}{\cosh2K-1}}.\label{eq:m_beq}\end{equation}

\bibliographystyle{apsrev}
\bibliography{/Users/fred/TeX/bib/Physik}

\end{document}